\documentclass{agujournal2019}
\usepackage{amsmath,amssymb,url} %
\usepackage{soul,textcomp}
\usepackage[normalem]{ulem}
\usepackage{wasysym,gensymb,csquotes}
\usepackage{rotating}

\ifdefined\pdfminorversion\pdfminorversion=7\fi

\journalname{JGR: Space Physics}

\makeatletter
\def\ps@headings{\def\@oddfoot{\centerline{\small\the\c@page}}\let\@evenfoot\@oddfoot
  \def\@oddhead{}\let\@evenhead\@oddhead}
\ps@headings
\AtBeginDocument{\nolinenumbers\let\linenumbers\relax}
\makeatother

\graphicspath{{./figs/}}

\def\re{\ensuremath{R_{\mathrm{E}}}}
\def\amin{\ensuremath{{\cal A}_{\mathrm{min}}}}
\def\amax{\ensuremath{{\cal A}_{\mathrm{max}}}}
\def\alfven{{Alfv\'{e}n}}

\definecolor{darkmagenta}{rgb}{0.8,0,0.8}
\definecolor{bluegrotto}{rgb}{.055, .525, .831}

\newcommand{\aguKeywords}[1]{\par\medskip\noindent{\bfseries Keywords:} #1\par}
\newcommand{\aguIndexTerms}[1]{\par\noindent{\bfseries Index terms:} #1\par}

\begin{document}

\title{A Survey of Electron Anisotropies in the Magnetosphere: Dependence on Solar Wind Pressure and $K_p$}

\authors{A. J. Hull\affil{1}, O. Agapitov\affil{1}, F. S. Mozer\affil{1}}
\affiliation{1}{Space Sciences Laboratory, University of California, Berkeley, Ca, USA.}

\correspondingauthor{O. Agapitov}{oleksiy.agapitov@gmail.com}

\begin{keypoints}
\item THEMIS statistics show that field-aligned and perpendicular electron anisotropies occupy distinct, energy-dependent regions of L-MLT space
\item Parallel (transverse) anisotropy regions expand (contract)  toward lower L and dusk (prenoon) during high solar wind dynamic pressure and Kp
\item Perpendicular anisotropy regions match quasi-parallel chorus sources; field-aligned regions match oblique chorus and TDS
\end{keypoints}

\begin{abstract}
We present a survey of electron distributions measured by THEMIS within Earth's equatorial magnetosphere at $L$ values $\le$ 14.
Particular attention is focused on the spatial distribution and occurrence of magnetic field-aligned and perpendicular anisotropies in electron distribution functions and how these depend on solar wind dynamic pressure and magnetospheric activity.
Using 161,300 3-s resolution electron measurements from THEMIS A and D, we find that field-aligned anisotropies are most prominent at low energies (30--200 eV) and occur over nearly all $L$ and MLT, with the largest values concentrated in the dayside/postnoon plasmasphere, while perpendicular anisotropies dominate at higher energies (1--30 keV) and are largest and most probable in a narrower band from $L \sim$ 5--10 with a pronounced dayside/prenoon bias.
A localized region near dawn at $4\lesssim L\lesssim 8$--9 is found to be typically devoid of field-aligned electrons over the entire 30--200~eV band.
Both anisotropy types shift toward lower $L$ shell and become more probable under elevated solar wind dynamic pressure and Kp, and their spatial and energy dependence is largely complementary: perpendicular anisotropies coincide spatially with regions of quasi-parallel chorus generation, while field-aligned anisotropies coincide with oblique chorus and time domain structure (TDS) generation regions.
In addition to being a useful diagnostic for electron population types, our results provide a useful basis for comparison with global distributions of wave modes which arise from these anisotropies, and offer an anisotropy-based observational framework for interpreting their dependence on magnetospheric activity.
\end{abstract}

\section*{Plain Language Summary}
Earth is surrounded by a magnetic bubble, the magnetosphere, filled with electrically charged particles. The electrons there do not move in random directions. In some regions they travel mostly along the magnetic field, like beads sliding along a wire; in others they circle around it. Which of these two behaviors dominates tells us where the electrons came from and how they were energized, and it also determines which kinds of natural radio waves that region can generate. Those waves matter, because they can accelerate electrons to energies that damage satellites, or scatter them into the upper atmosphere where they produce the aurora. Using measurements made by two of NASA's THEMIS spacecraft over fourteen months, we mapped where each electron motion type occurs, at three different energies ranges, and how those maps change when the solar wind pushes harder on the magnetosphere or magnetic activity increases. We find that the two motion types occupy almost entirely separate regions of space, that both patterns move closer to Earth during disturbed conditions, and that they line up with regions where different types of radio waves are known to be generated. The maps therefore show where the energy that drives those waves is stored.

\aguKeywords{electron pitch-angle anisotropy; Earth's equatorial magnetosphere and plasmapshere; field-aligned electrons; magnetospheric activity; whistler-mode chorus;
solar wind dynamic pressure}
\aguIndexTerms{2720 Energetic particles: trapped; 2731 Magnetosphere: outer;
2740 Magnetospheric configuration and dynamics; 2772 Plasma waves and instabilities;
2784 Solar wind/magnetosphere interactions}

\section{Introduction}

The electron distribution functions observed in Earth's equatorial magnetosphere are far from Maxwellian.
They are composed of multiple populations, of solar wind and/or ionospheric origin, that have been transported therein and energized through a variety of channels.
The relative contributions of these populations vary systematically with location and with energy, and they manifest themselves on the measured distribution functions as pitch-angle anisotropies that may be preferentially magnetic field-aligned (FAL), preferentially transverse (perpendicular) to the magnetic field, or effectively isotropic.
Two consequences follow, and together they motivate the present study.
First, anisotropy is a kinetic-level diagnostic of the plasma itself: the sign, size, and energy dependence of the anisotropy indicate which source populations have access to a given region of the magnetosphere and where the associated energization and transport processes are operative and effective in producing the observed signatures.
Second, when sufficiently large, distribution function anisotropies constitute a reservoir of free energy for the growth of plasma waves.
These waves in turn scatter, energize, and heat the plasma through wave-particle interactions, with important ramifications for particle sources and losses and, ultimately, for the dynamics of the magnetosphere and its coupling to the ionosphere.
A comprehensive determination of the size, occurrence probability, and spatial distribution of both FAL and transverse anisotropies therefore serves a dual purpose: it maps the domains occupied by distinguishable classes of electrons, and it provides an observational basis for interpreting where and under what conditions the wave modes driven by those anisotropies are expected to grow.

Anisotropies in magnetospheric electron distributions have been addressed in a number of previous studies \cite{abeletal:2002,hadaetal:1981,klumparetal:1988,moore+arnoldy:1982,arnoldy:1986,artemyevetal:2014,lietal:2010,walshetal:2011,walshetal:2013,hulletal:2021,hulletal:2020b,dentonetal:2017,mozeretal:2017}.
Preferential FAL anisotropies generally arise from the presence of a population of FAL electrons, which may appear as counterstreaming (bidirectional) or as unidirectional.
Counterstreaming FAL electrons with a preferential skew, or unidirectional FAL electrons, are indicative of the FAL currents that couple the magnetosphere to the ionosphere.
FAL electrons are observed on both the dayside and the nightside magnetosphere.
On the dayside, they have been reported in the inner magnetosphere \cite{dentonetal:2017}, in the outer magnetosphere \cite{mozeretal:2017}, and in the vicinity of the magnetopause \cite{oierosetetal:2015,mozeretal:2016}.
The THEMIS-based study of \citeA{mozeretal:2017} showed that counterstreaming FAL electrons occur over a large extent of the dayside outer magnetosphere ($L \ge 5$) within a given orbital transect, often spanning a broad range of energies from a few eV to hundreds of eV.
These electrons can be extremely field-aligned, with parallel-to-perpendicular flux ratios generally above 10 and at times exceeding 100.
When low-energy electrons at energies $\le$~25~eV are present, they generally provide the dominant contribution to the density, with typical values of $\sim$0.5--1~cm$^{-3}$ but at times an order of magnitude larger.
\citeA{mozeretal:2017} further reported that the occurrence probability of events with anisotropies above 100 is seasonally dependent, being $\sim$50\% in June and $\sim$10\% in September, and attributed these electrons to a dayside ionospheric source associated with the tilt of the northern cusp toward the Sun during summer.
Using measurements from the Van Allen Probes and LANL satellites, \citeA{dentonetal:2017} showed that FAL electrons at tens of eV energies are observed at all $L$ between 1 and 6.6 in the dayside inner magnetosphere during both quiet and active times, with the most intense fluxes occurring at $L$ between 1 and 3 and at $L \gtrsim 5$.
The average FAL fluxes at a given $L$ were elevated during low solar wind driving (as measured by $-V_{\rm SW}B_z < -1$) relative to periods of elevated driving ($-V_{\rm SW}B_z > 1$), suggesting that the preferential dayside fluxes of low-energy, FAL electrons result from outflows energized by solar EUV flux incident on the sunlit magnetic footpoints in the ionosphere.

FAL electrons are also a persistent feature of the nightside magnetosphere. The early statistical study of \citeA{hadaetal:1981}, based on IMP~6 observations, found bidirectional electrons at energies from several hundreds of eV to several keV in the plasma sheet near the neutral sheet at distances of 8--30~\re, at a rate of 10\% of the $\sim$200~hour intervals examined. Later work indicates that FAL electrons at sub-keV energies are observed most of the time in the magnetotail at distances $> 15$~\re\ \cite{walshetal:2011,walshetal:2013}.
FAL electrons at energies $\le$ a few keV in the magnetotail have been associated with substorm injections, magnetic field dipolarizations, and bursty bulk flows \cite{moore+arnoldy:1982,arnoldy:1986,sergeevetal:2001,chastonetal:2012,chastonetal:2015b,artemyevetal:2014,hulletal:2020b}.
\citeA{artemyevetal:2014} showed that the FAL anisotropies of electrons at and below several keV near the equatorial magnetotail at distances of 5--20~\re\ depend on location.
At 12--20~\re, preferential parallel anisotropies near midnight extend over most energies up to several keV, whereas in the dawn and dusk flanks they are limited to energies $<$~1~keV.
Closer to Earth (5--12~\re), the parallel anisotropy energy range is restricted to colder electrons (200--300~eV), with a notable dawnside asymmetry. Those authors suggested that parallel anisotropies at energies less than a few keV within $|x| < 12$~\re\ are likely of ionospheric origin, with the enhanced flank anisotropies attributed to a greater degree of ionization in the dawnside and duskside ionospheres relative to their midnight counterpart.
However, more energetic FAL anisotropies (at energies $<$~20~keV) were interpreted as being produced during reconnection and/or dipolarization and subsequently transported earthward and to the flanks by convection.
FAL electrons at and below a few keV in both the outer and inner magnetosphere plasma sheet have also been linked to kinetic \alfven\ waves \cite{chastonetal:2012,chastonetal:2015b,hulletal:2020b}.
In particular, \citeA{hulletal:2020b} showed that fluxes of FAL electrons are strongly enhanced in the inner magnetosphere plasma sheet ($L \sim$~3--6.6) during active geomagnetic conditions, producing preferential FAL anisotropies ranging from 1.2 to $>$~2 at energies up to several keV, with the highest energies occurring at magnetic field dipolarizations.
Detailed comparisons revealed strong correlations between the FAL electron energies and energy fluxes and the kinetic/dispersive \alfven\ wave (KAW) energy densities and Poynting fluxes, suggesting that KAWs play an important role in energizing these electrons along the field line.

Electron distributions with preferential perpendicular anisotropies, in contrast, are generally observed closer to Earth (e.g., $L \lesssim 9$).
Such distributions are commonly understood to develop as plasma sheet electrons convect earthward and are adiabatically energized in the increasing magnetic field, with pitch-angle scattering near the loss cone acting to further sharpen the transverse peak, giving rise to the ``pancake'' and butterfly distributions long recognized in the trough and inner magnetosphere \cite{wrennetal:1979,lyonsetal:1972,korthetal:1999}.
The global organization of these anisotropies is exemplified by the statistical study of \citeA{lietal:2010}, who examined electrons at energies between 500~eV and 200~keV observed by THEMIS near the equatorial outer magnetosphere (5--10~\re).
Their results show that during quiet substorm periods the perpendicular anisotropies of 0.5--30~keV electrons have a dayside/prenoon bias peaking at 7~$< L <$~9, with perpendicular anisotropies confined to $L <$~8 on the nightside albeit at lower levels.
During moderate to active conditions, notable increases are observed from midnight into the dawn and noon sectors, with exceptions in a narrow energy range about a few keV.
\citeA{bortniketal:2007} reported the average distribution of perpendicular electron fluxes at 0.213--16.5~keV as a function of $AE$ index observed by CRRES in the inner magnetosphere ($L < 7$).
They showed that perpendicular electron fluxes are significantly enhanced on the nightside relative to the dayside, with penetration to lower $L$ during substorm-active periods in association with enhanced convection \cite{korthetal:1999}, in contrast to the more uniform coverage seen during quiet periods.
A preferential nightside asymmetry was also visible in the noon and midnight distributions of fluxes as a function of pitch angle and $L$ for electrons from a few tens of eV to 1~keV reported by \citeA{dentonetal:2017}.
Those authors additionally showed that perpendicular fluxes for $\sim$1~keV electrons on the nightside were significantly enhanced during elevated solar wind driving relative to low driving, while the dayside fluxes at the same energy showed the opposite behavior, being enhanced at 4~$< L <$~7 during low solar wind driving.

It is well established that anisotropies in electron distribution functions drive the generation and evolution of a variety of plasma waves in the magnetosphere, and that the two anisotropy classes described above feed distinct wave populations.
Transverse anisotropies of $\sim$1--100~keV electrons drive cyclotron-resonant growth of quasi-parallel whistler-mode chorus near the geomagnetic equator \cite{kennel+petschek:1966,helliwell:1967,gary+wang:1996,Omura2009}, which is observed predominantly outside the plasmasphere with a dawnside/prenoon bias and a strong substorm dependence \cite{tsurutani+smith:1974,meredithetal:2001,Li2012,Meredith2012,Agapitov2013,agapitovetal:2018}.
Chorus is a major driver of local acceleration of outer radiation belt electrons to relativistic energies \cite{Summers2002,Horne2005,Thorne2013,Li2015} and, through pitch-angle scattering, of the diffuse auroral precipitation of plasma sheet electrons \cite{thorne:2010,Ni2011a}.
Transverse anisotropies also support whistler-mode hiss within the plasmasphere and in plasmaspheric plumes, which is a major cause of radiation belt electron loss \cite{lyonsetal:1972,meredithetal:2004,summersetal:2008}, and which is at least partly supplied by chorus that propagates into the plasmasphere \cite{bortniketal:2008,hartleyetal:2019}.
In contrast, FAL anisotropies favor a different set of modes.
A field-aligned plateau or weak beam at low parallel velocities suppresses Landau damping and permits whistler-mode growth at large wave-normal angles, so that low-energy field-aligned populations are implicated in the generation of very oblique, lower-band chorus \cite{Mourenas2015,artemyev:2016,Artemyev2016,Li2016,Li2016b,agapitovetal:2015,Agapitov2016}.
FAL electrons in the $\sim$50--1000~eV range likewise drive the linear electron-acoustic instability responsible for the broadband electrostatic solitary structures--commonly referred to as Time Domain Structures (TDS)--that occur near the geomagnetic equator and in the auroral zone \cite{mozeretal:2015,malaspinaetal:2015,vaskoetal:2017,agapitovetal:2018b}.
All of these waves play an essential role in particle energization and/or scattering, and in driving electron precipitation into the ionosphere; through these processes, they act as important mediators of magnetospheric energy conversion and transport.
Understanding the properties of electron anisotropies in the magnetosphere is therefore essential, since those anisotropies are the direct tracers of wave growth and of the energization processes that sustain it.

Although significant progress has been made, a global, kinetic-level picture of how the two anisotropy classes are organized in $L$, MLT, magnetic latitude, and energy--and of how that organization responds to solar wind driving and magnetospheric activity--remains incomplete.
Existing surveys have generally been restricted in energy coverage (e.g., $\ge$~500~eV in \citeA{lietal:2010}; $\le$~1~keV in the pitch-angle comparisons of \citeA{dentonetal:2017}), in radial range (5--10~\re\ in \citeA{lietal:2010}; $L \le 6.6$ in \citeA{dentonetal:2017}; $L < 7$ in \citeA{bortniketal:2007}), or have characterized anisotropy through temperature moments.
The latter suffers a particular limitation in that moments average over the disparate populations that coexist within a single distribution: a cold, intensely field-aligned beam superposed on a hotter, transversely anisotropic trapped population can yield a temperature ratio near unity even though both anisotropy classes--and hence the free energy for two distinct wave modes--are simultaneously present.
In addition, the relative roles of external driving (solar wind dynamic pressure) and internal activity ($Kp$) in controlling where each anisotropy class resides have not been systematically separated over a radial range that spans the plasmasphere, the inner magnetosphere, and the outer magnetosphere together.

In this paper we address these issues through case and statistical analysis of anisotropies in electron distribution functions within the near-Earth magnetosphere at $L \le 14$, using observations from the THEMIS~A and D spacecraft \cite{angelopoulos:2008,mcfaddenetal:2008} sampled from 21 June 2016 to 31 August 2017.
The data span the near-equatorial outer magnetosphere, the inner magnetosphere, and the plasmasphere.
Rather than relying on temperature moments, we quantify anisotropy at the kinetic level using the extrema of the parallel-to-perpendicular differential energy flux ratio evaluated across the measured energy range, $\amin = \min(J_{e\parallel}/J_{e\perp})$ and $\amax = \max(J_{e\parallel}/J_{e\perp})$, which respectively isolate the peak transverse and the peak FAL anisotropy exhibited by each distribution.
A database of 161,300 3-s spin-resolution measurements is sorted into 1~hr MLT by 1~$L$ bins within three energy ranges--low (30--200~eV), middle (0.2--1~keV), and high (1--30~keV)--chosen to encompass the disparate anisotropic populations, and is further sorted by solar wind dynamic pressure $P_{\rm SW}$ and by $Kp$.

The resulting picture is one in which the two anisotropy classes occupy largely complementary, energy-dependent domains.
FAL anisotropies are most prominent at low energies and occur over nearly all $L$ and MLT, with the largest values concentrated in the dayside/postnoon plasmasphere and an asymmetric horseshoe-like pattern of high occurrence centered on the premidnight sector of the outer magnetosphere, together with a pronounced occurrence dropout on the dawnside at $L \sim$~5--10.
Perpendicular anisotropies dominate at higher energies and are largest and most probable within a narrower band from $L \sim$~5--10 with a pronounced dayside/prenoon bias.
Both anisotropy types shift toward lower $L$ shell and become more probable under elevated $P_{\rm SW}$ and $Kp$. Comparison with the statistical distributions of whistler-mode chorus and hiss shows that the transverse-anisotropy region coincides spatially with the quasi-parallel chorus source region, while the FAL anisotropy region coincides with the regions favorable to oblique chorus and TDS generation. In this way, the study elucidates the domains of distinguishable classes of magnetospheric electrons and, because anisotropies are the free-energy source for these modes, provides a useful basis for comparison with statistical wave surveys and an anisotropy-based observational framework for assessing wave growth and its consequences.

The paper is organized in the following manner.
Section~\ref{sec:instrument} describes the instrumentation, experimental data set, event selection, and anisotropy diagnostics. Section~\ref{sec:event} presents a case event that illustrates the typical characteristics of electron anisotropies associated with the different populations observed in the magnetosphere.
Statistical distributions and occurrences of parallel and perpendicular electron anisotropies, and their dependence on solar wind dynamic pressure and $Kp$, are given in Section~\ref{sec:stats}.
The discussion is given in Section~\ref{sec:discussion}, where we compare the anisotropy distributions with the distributions of whistler-mode chorus and hiss, and discuss the implications of the anisotropy distributions for the causes and effects of other wave modes in the magnetosphere.
Finally, the summary and conclusions are presented in Section~\ref{sec:conclusions}.

\section{Instrumentation, Experimental Data Set, Event Selection, and Anisotropy Diagnostics}
\label{sec:instrument}

This study is primarily based on electron measurements from the ElectroStatic Analyzer (ESA) instrument onboard the THEMIS A and D spacecraft \cite{angelopoulos:2008,mcfaddenetal:2008}.
A pair of ESAs are onboard each spacecraft to measure electrons and ions at energies $<$~30 keV.
The anisotropy analysis presented here uses full three-dimensional electron distributions sampled at spin resolution (3~s) with a cadence of $\sim$6.7~min in slow survey mode and $\sim$1.68~min in fast survey mode.
Electron moments used in the study are derived from the electron distribution functions after correcting for the floating spacecraft potential provided by the Electric Field Investigation (EFI) \cite{bonnelletal:2008}.
Spin period magnetic field measurements from the FluxGate Magnetometer (FGM) \cite{austeretal:2008} were also used to provide context and to determine perpendicular and parallel energy fluxes used in quantifying electron anisotropies.

To select data corresponding to magnetospheric electrons, we required the temperature of hot electrons ($\ge$ 25 eV energies above the spacecraft potential) to be larger than 100 eV.
Given that magnetosheath electrons typically have post-shocked values of up to several tens of eV, this criteria is quite effective in discriminating intervals in the magnetosheath plasma from those in the magnetosphere \cite{walshetal:2020}.

NASA OMNIWeb solar wind parameters (e.g., dynamic pressure) mapped to the bow shock nose and magnetospheric activity measures (e.g., $Kp$) were also used in this study to examine dependencies on solar wind driving conditions and magnetospheric activity.
Following common practice, no additional propagation delay beyond the bow shock nose was applied; the residual $\sim$10--20~min response time of the inner magnetosphere is short compared with the 3~hr cadence of $Kp$ and is not expected to affect the statistical sorting used below. 

Throughout this study, anisotropy is quantified at the level of the distribution function using flux ratios evaluated in three fixed pitch-angle sectors, rather than through temperature moments. The anisotropy at a given energy is
\begin{linenomath*}
\begin{equation}
{\cal A} = \frac{J_{e\parallel} + J_{e-\parallel}}{2\,J_{e\perp}},
\label{eq:aniso}
\end{equation}
\end{linenomath*}

where $J_{e\parallel}$, $J_{e-\parallel}$, and $J_{e\perp}$ are the differential energy fluxes averaged over pitch angles $\alpha = 0^{\circ}$--$20^{\circ}$, $160^{\circ}$--$180^{\circ}$, and $80^{\circ}$--$100^{\circ}$, respectively. The two diagnostics used throughout the statistical analysis are the extrema of ${\cal A}$ over a specified energy band,
\begin{linenomath*}
\begin{equation}
\amax \equiv \max({\cal A}), \qquad \amin \equiv \min({\cal A}),
\label{eq:aminmax}
\end{equation}
\end{linenomath*}
so that $\amax$ measures the largest FAL anisotropy and $\amin^{-1}$ the largest transverse anisotropy exhibited anywhere within the band. Values $\amax > 1$ indicate that a preferential FAL anisotropy is present somewhere in the band, while $\amax < 1$ indicates that only transverse anisotropies occur over the full band and therefore represents the \emph{smallest} transverse anisotropy present. Conversely, $\amin < 1$ indicates that a transverse anisotropy is present somewhere in the band, while $\amin > 1$ represents the smallest FAL anisotropy present. 
Utilizing both of these parameters in anisotropy diagnostics is key in that, because the two extrema are evaluated independently, they retain the ability to distinguish whether a single distribution is field-aligned at one energy and transverse at another, or if it is transverse (i.e., $\amin \lesssim 1$ and $\amax \lesssim 1$) or field-aligned (i.e., $\amin \gtrsim 1$ and $\amax \gtrsim 1$) across the entire energy band. 
We take advantage of this pairing to identify different electron populations as well as distinct regions in the geophysical maps presented below.

Two instrumental effects should be borne in mind in what follows. First, penetrating outer radiation belt electrons produce a spurious field-aligned signature in the 0.1--1~keV and 1--30~keV ranges within a narrow band near $L\sim$~4; this feature is identified where it occurs in the figures below and is excluded from interpretation. Second, the low-energy band lies close to the spacecraft potential, so that the anisotropies derived there, particularly the large plasmaspheric values, are sensitive to the choice of energy floor.

\section{Electron Anisotropy Event Case Study}
\label{sec:event}

\begin{figure}
\vspace{-30pt}
\center{\includegraphics[width=5.5in]{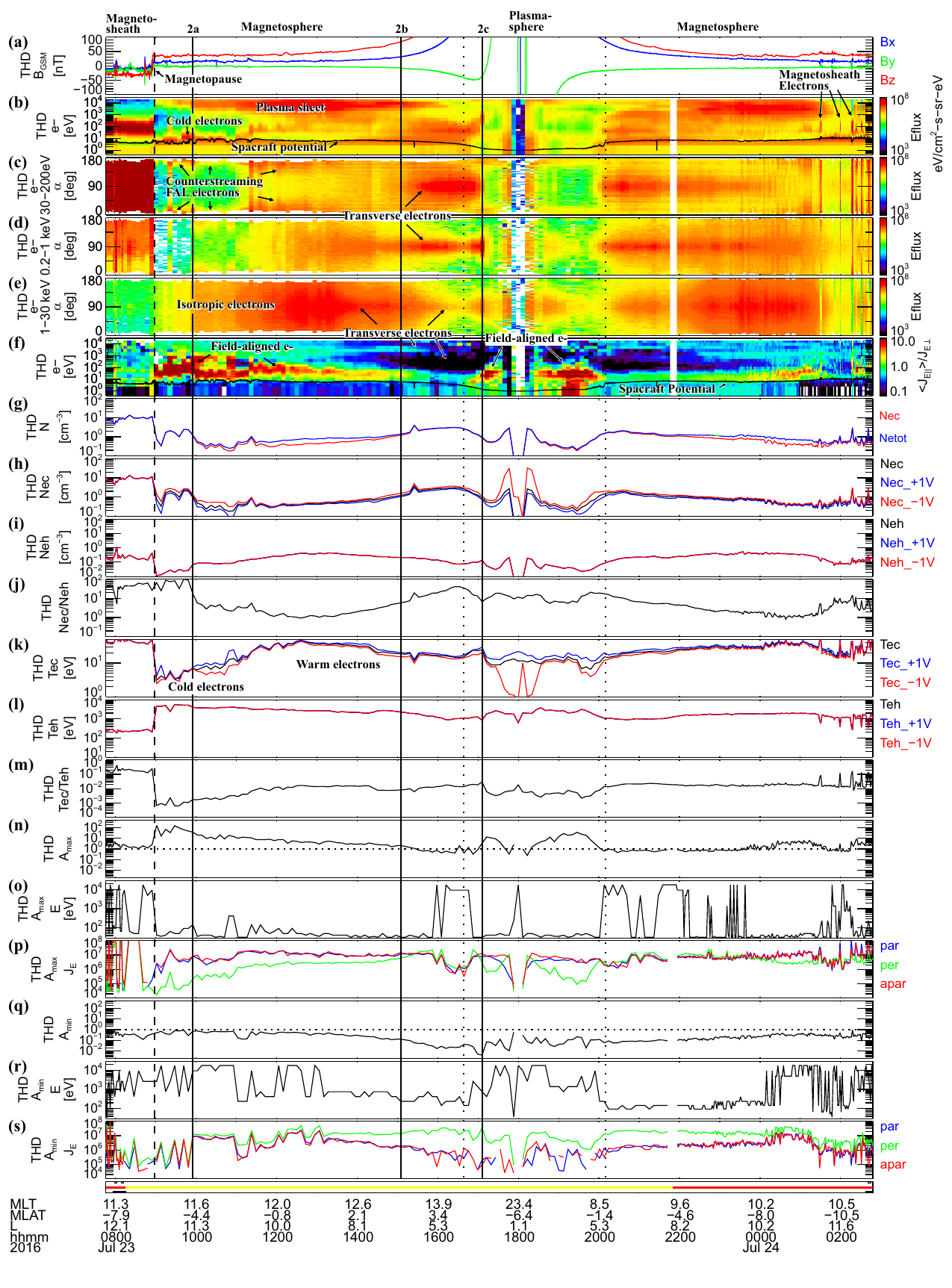}}
\vspace{-30pt}
\caption{(a) Magnetic field, (b) electron energy flux energy spectra, energy flux pitch-angle spectra for (c) 30-200 eV, (d) 0.2-1 keV, and (e) 1-30 keV electrons, (f) electron anisotropy spectra, (g) densities, densities of (h) cold and (i) hot electrons, (j) cold-hot electron density ratio, temperatures of (k) cold and (l) hot electrons, (m) cold-hot electron density ratio, (n) maximum electron anisotropy (${\cal A}_{\rm max} = {\rm max}(J_{e\parallel}/J_{e\perp})$), (o) energy at ${\cal A}_{\rm max}$, (p) parallel (blue), perpendicular (green), and antiparallel (red) differential energy fluxes at ${\cal A}_{\rm max}$, (q) minimum electron anisotropy ${\cal A}_{\rm min}$, (r) energy at ${\cal A}_{\rm min}$, and (s) parallel, perpendicular, and antiparallel differential energy fluxes at ${\cal A}_{\rm min}$. The solid curve in (b) and (f) indicates the spacecraft potential. The yellow and red in the bottom bar indicate slow and fast survey mode plasma data sampling rates.}
\label{fig:event1}
\end{figure}

To illustrate the properties of electron anisotropies in the magnetosphere, Figure \ref{fig:event1} shows an inbound to outbound crossing of the magnetosphere and plasmasphere regions by the THEMIS D spacecraft on 23--24 July 2016.
The event occurred under relatively quiet magnetospheric conditions (median $Kp = 1^+$) and low solar wind dynamic pressure (median $P_{\rm SW}$~=~0.85~nPa).
The spacecraft was initially in the magnetosheath, crossed the magnetopause into the magnetosphere at $\sim$0900~UT on 23 July 2016 near 11.4 MLT and $L\sim 11.7$ (dashed vertical line), moved inward to the plasmasphere at $\sim$1630~UT near $L\sim 4.4$ (first dotted line), and reached perigee at $\sim$1800 UT. 
It then exited the plasmasphere on the outbound leg at $\sim$2008~UT near $L\sim 5.6$ (second dotted line) and remained in the magnetosphere until it made multiple transient magnetosheath crossings starting at $\sim$0128 UT on 24 July 2016.
Key regions are indicated at the top of Figure \ref{fig:event1}.

Entry into the magnetosphere is marked by the transition from highly variable, lower-amplitude magnetosheath fields to a more dipolar configuration with a dominant $B_{Z}$ component (Figure \ref{fig:event1}a) and by the appearance of plasma sheet electrons at energies $\gtrsim$~1 keV in Figure \ref{fig:event1}b.
This population spans from the magnetopause to the plasmasphere on both transects and extends to lower energies closer to Earth.
On the outbound leg the plasma sheet electrons near the inner boundary exhibit the characteristic energy dispersion arising from the energy dependence of accessible drift paths of electrons originating in the magnetotail, with lower energies seen first \cite<e.g.,>[]{thomsenetal:2002}.
On the inbound leg, the behavior is more complex, likely owing to a mixture of injected populations arising from variable convection conditions.
The plasma sheet electrons are tenuous and hot, with densities and temperatures of 0.01--0.08 cm$^{-3}$ and 5 keV near the magnetopause (Figures \ref{fig:event1}i and \ref{fig:event1}l), and $\sim$0.5 cm$^{-3}$ and $\sim$1 keV deeper in the magnetosphere.
Also apparent in Figure \ref{fig:event1}b is a dense, cold, low-energy population near the magnetopause (from $\sim$0900--1010~UT), with densities peaking at $\sim$2--3 cm$^{-3}$ and temperatures of a few eV (Figures \ref{fig:event1}h and \ref{fig:event1}k).
These electrons dominate the density, as evidenced by the matching cold (red) and total (blue) traces in Figure \ref{fig:event1}g and by cold-to-hot density ratios reaching $\sim$100 near the magnetopause, and by cold-to-hot temperature ratios of $\sim$0.001 (Figure \ref{fig:event1}m).
The cold and hot integration ranges are 0--25~eV and 25--30,000~eV above the spacecraft potential, respectively.
As the spacecraft moves inward, this population becomes warmer (20--50 eV), decreasing in density before increasing again prior to plasmasphere entry.

The pitch-angle behavior of these populations is shown in Figures \ref{fig:event1}c--\ref{fig:event1}e, which give differential energy fluxes versus pitch angle for the three energy ranges, and in Figure \ref{fig:event1}f, which shows the anisotropy ${\cal A}$ of Equation~\ref{eq:aniso} as a function of energy and time.
In Figure \ref{fig:event1}f, preferential parallel anisotropies appear yellow to red, preferential perpendicular cyan to black, and isotropic green.

Figure \ref{fig:event1}f shows that different magnetospheric populations are distinguishable by their anisotropy characteristics.
On the inbound leg the plasma sheet electrons are on average isotropic near the magnetopause (green at $\gtrsim$1~keV from $\sim$0900--1300 UT), consistent with the uniform pitch-angle distribution in Figure \ref{fig:event1}e, whereas closer to Earth they develop a preferential perpendicular anisotropy (black to dark blue above a few tens of eV), appearing as a broad flux enhancement near 90\textdegree\ with symmetrically reduced fluxes near 0\textdegree\ and 180\textdegree.
A similar behavior occurs on the outbound leg, where the anisotropy transitions from strongly perpendicular at low $L$ to weakly perpendicular near the magnetopause.

In contrast, the low energy electrons often exhibit a preferential FAL anisotropy over extended regions outside the plasmasphere, indicated by the yellow to red signatures just above the spacecraft potential in Figure \ref{fig:event1}f and arising from the nearly symmetric counterstreaming character evident in Figures \ref{fig:event1}c and \ref{fig:event1}d.
These FAL anisotropies occupy a much larger region on the inbound leg (0900--1520 UT; 11.4--13.3 MLT, $L\sim$~11.75--6.44, $|\Delta L|\sim$5.31) than on the outbound leg (2219--0128 UT; 8.5--11.3 MLT, $L\sim$~8.55--11.27, $|\Delta L|\sim$2.72), an asymmetry indicative of a geophysical location dependence explored statistically in Section \ref{sec:stats}.
The FAL electrons span the broadest energy range and have the largest parallel anisotropies near the magnetopause--from just above the spacecraft potential to several hundred eV on the inbound transect, but limited to $\lesssim$100 eV on the outbound one--and both the energy extent and the anisotropy decrease with decreasing $L$ before the distributions become isotropic or weakly transverse ahead of the plasmasphere.
Large parallel anisotropies due to low energy FAL electrons are also observed within the plasmasphere itself, at energies from just above the spacecraft potential up to $\sim$100 eV.
Opposite to the behavior outside, these are confined closer to Earth on the inbound transect (14.9--18.1 MLT, $L\sim$~3.8--1.7) than on the outbound one (4.7--8.6 MLT, $L\sim$~1.9--5.6), reflecting the asymmetric shape of the plasmasphere in which they are confined.

The diagnostics defined in Equation~\ref{eq:aminmax} quantify this behavior.
Figure \ref{fig:event1}q shows that $\amin < 1$ almost everywhere, indicating that transverse anisotropies are typically present.
Near the magnetopause on both transects the peak transverse anisotropies are weak to moderate ($\amin\sim$0.7--0.9 and 0.2--0.3), and Figure \ref{fig:event1}r shows that they occur at 1--30 keV, i.e., in the energetic plasma sheet population.
At lower $L$ outside the plasmasphere the peak values are larger ($\amin\sim$0.03--0.1) and occur at 0.1--1 keV, in association with warmer plasma sheet constituents; large peak transverse anisotropies also occur within the plasmasphere at 1--30 keV.
Figure \ref{fig:event1}s shows that the increase in peak transverse anisotropy with decreasing $L$ is due to a symmetric reduction of the parallel and antiparallel fluxes accompanied by an enhancement of the perpendicular component.

Figures \ref{fig:event1}n and \ref{fig:event1}o show that the largest peak parallel anisotropies occur closest to the magnetopause on the inbound transect, with $\amax\sim$16--150 (median 42) at energies of 26--145~eV (median 51~eV).
Both $\amax$ and its associated energy decrease with decreasing $L$ until $\sim$1520~UT, beyond which the low energy electrons no longer exhibit parallel anisotropies.
Figure \ref{fig:event1}p indicates that this reduction is attributed to a relative enhancement of the transverse energy flux accompanied by a slight reduction of the field-aligned components with decreasing $L$, rather than to the disappearance of the field-aligned population alone.
The outbound transect shows the same trend at reduced amplitude ($\amax\sim$2.0--6.3, median 4.1, at 32--51~eV), with the transition to isotropy occurring at $\sim$2342~UT.
Within the plasmasphere, $\amax$ reaches $\sim$10--40 at energies of $\sim$40--50~eV.

\begin{figure}
\vspace{-8pt}
\center{\includegraphics[width=6.5in]{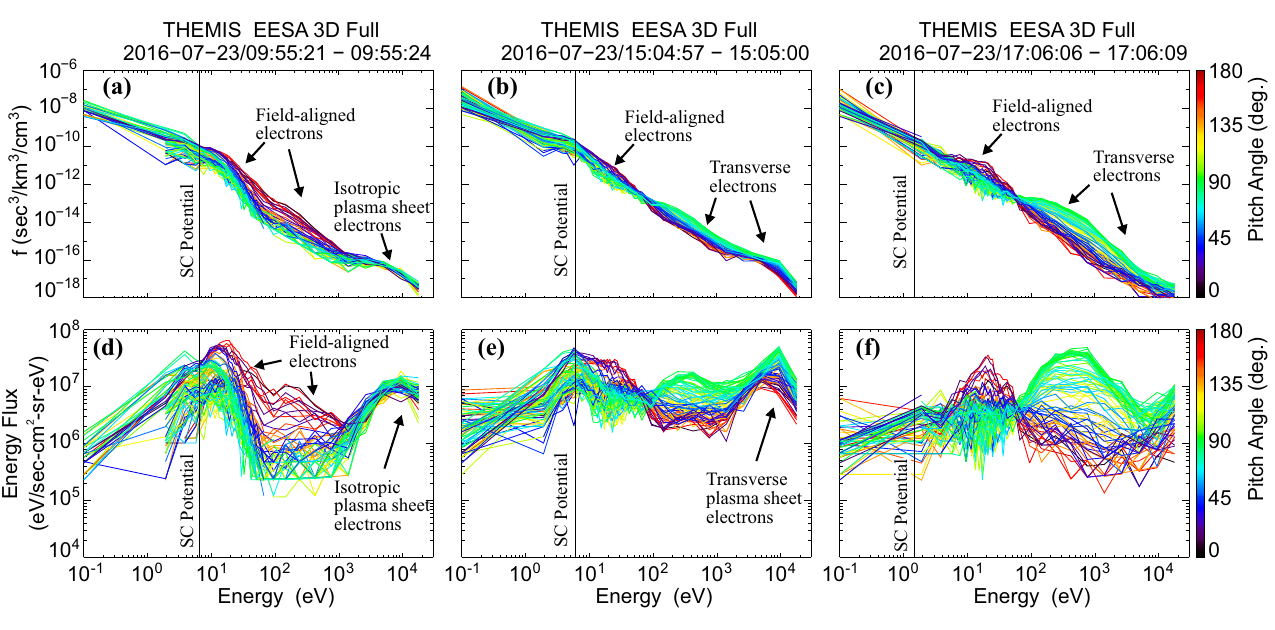}}
\vspace{-8pt}
\caption{Examples 3s resolution electron distributions from the July 23, 2016 event sampled at times indicated by the vertical lines in Figure \ref{fig:event1}.  Shown are (a-c) electron phase space density and (d-f) electron differential energy flux spectra as a function of energy for different pitch-angles.}
\label{fig:edist_event1}
\end{figure}

Figures \ref{fig:edist_event1}a--\ref{fig:edist_event1}c and \ref{fig:edist_event1}d--\ref{fig:edist_event1}f show example 3~s phase space density and differential energy flux spectra, color coded by pitch angle, sampled at the times indicated by the labeled vertical solid lines in Figure \ref{fig:event1}.
The first example, sampled near the magnetopause, is composed of highly field-aligned counterstreaming electrons spanning from a few eV above the spacecraft potential--where they dominate the density--up to $\sim$1~keV, with characteristics similar to the extremely field-aligned electrons reported by \citeA{mozeretal:2017}; here $\amax = 37$ at 145~eV.
A plasma sheet population also exists at $\gtrsim$1~keV, which is roughly isotropic,  except near the distribution peak at 9.3 keV, where $\amin^{-1}= 1.4$.
The second example, sampled deeper in the magnetosphere, resolves three populations: a counterstreaming FAL population from the spacecraft potential to $\sim$70 eV with a reduced $\amax = 2$ at 32~eV; a warm, transversely anisotropic plasma sheet population from 100 eV--1~keV; and an energetic, transversely anisotropic plasma sheet population above 1~keV.
The warm population carries the largest transverse anisotropy, $\amin^{-1}= 7.8$ at 240 eV.
The third example, sampled in the plasmasphere, exhibits counterstreaming FAL electrons from 5 to 60~eV with $\amax\sim 10$ at 25~eV, and above 60~eV a large transverse anisotropy peaking at $\amin^{-1}= 250$ at $\sim$760 eV.

Taken together, the case event demonstrates that the three populations resolved here--low energy FAL electrons, a warm transverse population, and hot plasma sheet electrons--carry distinguishable and spatially organized anisotropy signatures, motivating the statistical survey that follows.

\section{Statistics}
\label{sec:stats}

\subsection{Distribution of Anisotropies}
\label{sec:aniso_dist}

 \begin{sidewaysfigure}
 \center{\includegraphics[width=9in]{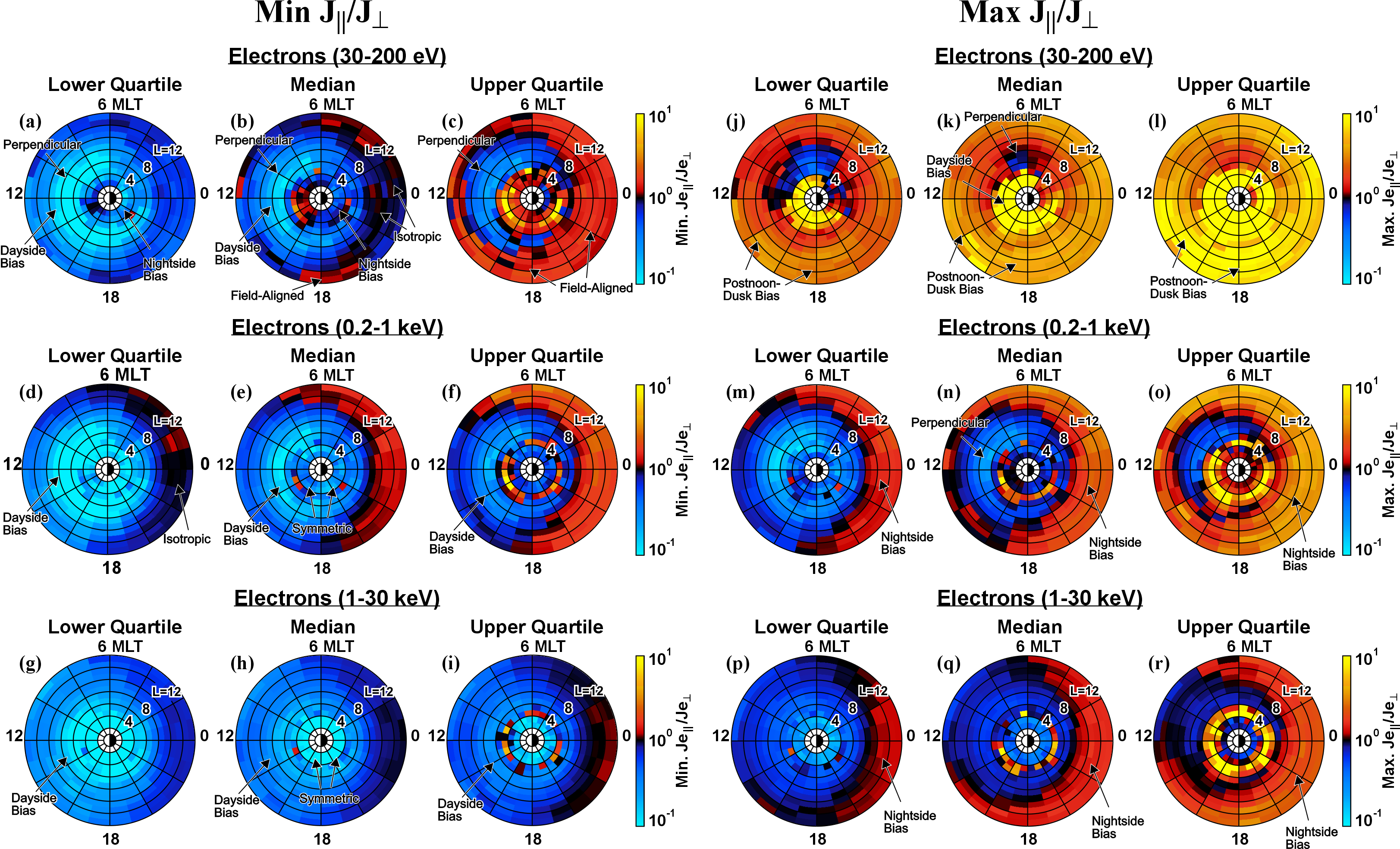}} 
\caption{(a, d, and g) Lower quartile, (b, e, and h) median, and (c, f, and i) upper quartile distributions of minimum electron flux ratios $J_{e\parallel}/J_{e\perp}$ in the (a-c) 30 $<$ E $<$ 200 eV, (d-f) 0.2 $<$ E $<$ 1 keV, and (g-i) 1 $<$ E $<$ 30 keV energy ranges, (j, m, and p) lower quartile, (k, n, and q) median, and (l, o, and r) upper quartile distributions of maximum $J_{e\parallel}/J_{e\perp}$ in the (j-l) 30 $<$ E $<$ 200 eV, (m-o) 0.2 $<$ E $<$ 1 keV, and (p-r) 1 $<$ E $<$ 30 keV energy ranges.}
 \label{fig:minmaxanisotropy_lo-high_e}
 \end{sidewaysfigure}

To assess the global properties of electron anisotropies, we compiled a statistical database of 161,300 3~s spin resolution electron measurements from THEMIS A and D within Earth's magnetosphere, sorted them into 1~hr MLT by 1~$L$ shell bins, and computed the lower quartile, median, and upper quartile values of \amin\ and \amax\ within the low (30--200~eV), middle (0.2--1~keV), and high (1--30~keV) energy ranges.
The energy ranges were chosen to encompass the disparate anisotropic populations revealed in Figures \ref{fig:event1} and \ref{fig:edist_event1}.
Note, the two anisotropy types are not mutually exclusive within a given MLT-$L$ shell bin for a specified energy band, since each bin aggregates distributions that may have contributions from different populations at different energies (and conditions) within the band.
Figure \ref{fig:minmaxanisotropy_lo-high_e} presents the resulting synoptic maps: panels a--i give \amin\ (a diagnostic for the peak transverse anisotropy) and panels j--r give \amax\ (a diagnostic for the peak FAL anisotropy), in each case for the low, middle, and high energy ranges in successive rows.
Blue to cyan indicates preferential perpendicular anisotropies, red to yellow preferential parallel anisotropies, and black roughly isotropic flux ratios.

\subsubsection{Transverse anisotropies}

Figures \ref{fig:minmaxanisotropy_lo-high_e}a--\ref{fig:minmaxanisotropy_lo-high_e}i show that large transverse anisotropies occur in all three energy ranges over a broad extent of the inner and outer magnetosphere, and that each energy range exhibits a dayside bias.
In the 30--200~eV range, the bias is roughly symmetric about noon in the lower quartile and median maps but skews to the prenoon sector in the upper quartile map.
The largest transverse anisotropies occur on the dayside at $L\sim$5--9, 
weakening beyond $L>$9.
On the nightside, the transverse anisotropies are reduced and confined closer to Earth.
Beyond $L\gtrsim$8, the distributions are typically near-isotropic, with the quartile spread ranging from moderately transverse ($\amin\sim$0.51--0.66) to preferentially FAL ($\amin\sim$1.39--1.66), and near dusk and dawn at $L\gtrsim$12 only FAL anisotropies are typically exhibited.
The median map also shows a nightside/morning bias at $L \lesssim$~4--5, owing to the presence of FAL anisotropies on the dayside at those distances.
It is important to emphasize that preferential FAL values appearing in the \amin\ maps arise from distributions that exhibit \emph{no} transverse anisotropy anywhere in the band, and therefore represent parallel anisotropy floor levels.

In the 0.2--1~keV range, the dayside bias becomes more symmetric about noon at $L>$4 and the regions of largest transverse anisotropy broaden, extending to both higher and lower $L$ than in the low energy range.
Peak values at $L\sim$5--9 are slightly elevated relative to the low energy range, and significant transverse anisotropies now also appear at $L\sim$2--4 on the dayside, where the low energy range exhibited mainly FAL anisotropies.
Correspondingly, the nightside bias at $L \lesssim$~4 present at low energies disappears, since dayside and nightside values become comparable.
Parallel anisotropy floor values are typically present at $L \gtrsim$~10 in the dawn, midnight, and dusk sectors.

In the 1--30~keV range. the dayside bias persists but is less pronounced, being clearest at $L>$~8.
The moderate to high transverse anisotropies now appear closer to Earth than in the other two ranges.
The highest values are within the plasmasphere (at $L\lesssim$~4), which are approximately symmetric about all MLTs (typically \amin\~0.1 on both the dayside and nightside).
Transverse anisotropies just outside the plasmasphere are typically more moderate, decreasing more gradually with distance on the dayside than on the nightside.
Contrasting the mid-energy range, weak transverse to isotropic values now occur at $L \gtrsim$~10 on the nightside and dawn-dusk flanks.  
The systematic inward migration and strengthening of the transverse-anisotropy domain with increasing energy is the central energy-dependent trend of the \amin\ maps.

\subsubsection{Field-aligned anisotropies}

Large preferential FAL anisotropies also occur in all three energy ranges (Figures \ref{fig:minmaxanisotropy_lo-high_e}j--\ref{fig:minmaxanisotropy_lo-high_e}r), but with the opposite energy dependence: they are largest and most pervasive in the low energy range but systematically weaken and spatially contract with increasing energy.

In the 30--200~eV range (Figures \ref{fig:minmaxanisotropy_lo-high_e}j--\ref{fig:minmaxanisotropy_lo-high_e}l), FAL anisotropies occur over a broad extent of the inner and outer magnetosphere.
A distinct region from $L=$~2 up to $\sim$4--6, encompassing the plasmasphere, has very large values (bright yellow) with a pronounced dayside/postnoon bias: $\amax$ is typically 18.8 (off-scale) on the dayside at $L\le$4 but 5.6 on the nightside in the median quartile distribution.
These values are due to FAL electrons at energies up to $\sim$100~eV, exemplified by the plasmaspheric distributions shown in Figures \ref{fig:edist_event1}c and \ref{fig:edist_event1}f.
This region has characteristics similar to the preferential dayside FAL fluxes of 33 eV electrons reported by \citeA{dentonetal:2017} and interpreted by those authors as ionospheric electrons energized by incident solar EUV flux in daylight.
At larger distances ($L>$6), \amax\ first drops abruptly (in the trough) and then increases with increasing $L$.
The increase is notably more rapid from noon through dusk and into premidnight, resulting in a preferential duskside bias with median values of 3.4--4.8 (a few exceptions reaching $\sim$6--10 at larger $L$) against dawnside values of 1.8--3.3.
The same dawn-dusk asymmetry is present in the lower and upper quartile maps.%

Contributing to this asymmetry, and of particular interest here, is a localized region on the dawnside in which \emph{only} transverse anisotropies are observed over the entire 30--200~eV band, indicated by the dark blue in Figures \ref{fig:minmaxanisotropy_lo-high_e}j and \ref{fig:minmaxanisotropy_lo-high_e}k, where $\amax\sim$0.67--0.96.
This region is typically, though not always, devoid of FAL electrons across the full band, indicated by the values $\amin \lesssim 1$ and $\amax \lesssim 1$ in the lower and median quartile maps, but not in the upper quartile map. 
The cause is not established here.
This may be a reflection of the drift paths that have access to this region, the inaction of the FAL electron source mechanism there, or both.

In the 0.2--1~keV range (Figures \ref{fig:minmaxanisotropy_lo-high_e}m--\ref{fig:minmaxanisotropy_lo-high_e}o), FAL anisotropies are generally restricted to larger $L$ and reduced in magnitude relative to the low energy range, and the pattern acquires a nightside bias.
This stems from the expansion of regions associated with only transverse anisotropies over the full band, which is most pronounced in the dayside/prenoon outer magnetosphere.
In the median map, FAL anisotropies span $L\gtrsim$~8 on the nightside, $L\gtrsim$~10 at dusk and dawn, and $L\gtrsim$~12 post-dawn, with $\amax\sim$1.2--3.7 constituting a slightly asymmetric horseshoe-shaped pattern centered about the morning sector.
Intermittent low-anisotropy regions also appear at noon and postnoon at $L\gtrsim$~11.
The lower quartile map shows a more limited domain composed of lower values. The upper quartile map shows a broader one, with FAL anisotropies spanning $L\sim$~2--14 from dusk to midnight, owing in part to inner magnetosphere ($L\lesssim$5--6) contributions absent or diminished in the other two quartiles, and extending continuously from dawn through noon into dusk at lower $L$.
Note, the red and yellow localized in the thin region at $L\sim$4--5 in the median and upper quartile maps are affected by penetrating radiation belt electrons as stated in Section \ref{sec:instrument} and are not interpreted further.

In the 1--30~keV range (Figures \ref{fig:minmaxanisotropy_lo-high_e}p--\ref{fig:minmaxanisotropy_lo-high_e}r), the FAL anistropy domain and amplitudes become further reduced, in association with the expansion of the isotropic and transverse regions.
The median map shows FAL anisotropies occupying a horseshoe-shaped region in the outer magnetosphere, roughly symmetric about 21 MLT, extending from postnoon through midnight into dawn at $L\gtrsim$~12, $L\gtrsim$~8, and $L\gtrsim$~10, respectively, with low values $\amax\sim$1.2--1.5.
This premidnight-centered horseshoe contrasts with the morning-centered pattern of the mid-energy range, indicating that more than one FAL electron source and/or energization mechanism contributes across the band sampled here.
The upper quartile map shows a larger horseshoe reaching $L\sim$7 at premidnight and $L\sim$11 at dawn and postnoon, with $\amax\sim$1.2--2.0.
The values in the narrow band at $L\sim$~4--5 are again affected by penetrating radiation belt electrons.

\subsection{Anisotropy Dependence on Solar Wind Dynamic Pressure and $Kp$}
\label{sec:aniso_psw_kp}

\begin{figure}
 \center{\includegraphics[width=3.5in]{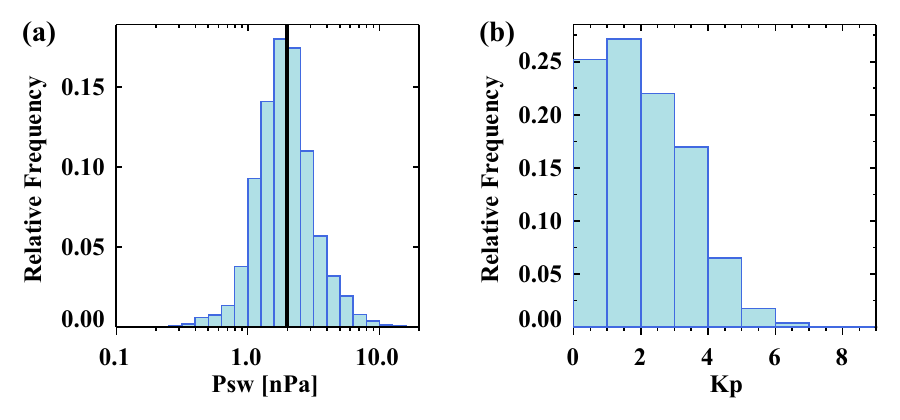}} 
\caption{Occurrence distributions of (a) solar wind dynamic pressure $P_{\rm SW}$ and (b) $Kp$.  The solid line in (a) indicates the median value.}
\label{fig:psw_kp_hist}
\end{figure}

To examine dependencies, the anisotropy data in each energy range were sorted by solar wind dynamic pressure $P_{\rm SW}$ and by $Kp$.
Figures \ref{fig:psw_kp_hist}a and \ref{fig:psw_kp_hist}b show the occurrence distributions of these parameters over the database.
Values of $P_{\rm SW}$ span 0.2--20.0 nPa, with a mean value of 2 nPa, which is used as the breakpoint between low and high pressure. 
Note that with a distribution mode of 1.78 nPa, the events occur more frequently below than above the breakpoint.
The $Kp$ occurrences peaks between 1 and 2, characteristic of quiet conditions.
In sorting by activity, we use $Kp\le 2^+$ ($Kp \le 2.33$) for quiet and $Kp \ge 3^-$ ($Kp\ge 2.67$) for active conditions, so that the two populations are contiguous in $Kp$.

Before describing the results, we mention here once two important points that apply to the comparisons below.
First, the $P_{\rm SW}$ and $Kp$ occurrence distributions indicate that the undifferentiated maps of \amin\ and \amax\ in all three energy ranges in Figure \ref{fig:minmaxanisotropy_lo-high_e} are reflective of quiet, low-pressure conditions and thus are found to be very similar to their corresponding low $P_{\rm SW}$ and low $Kp$ maps presented below.
In what follows, we describe only the differences that appear under elevated $P_{\rm SW}$ and $Kp$.
Second, $Kp$ and $P_{\rm SW}$ are moderately positively correlated over this epoch ($r=0.43\pm 0.02$ at 99\% confidence \cite{hulletal:2021}), so that the two parameters are not fully independent.
The differences between them reported below are correspondingly the differences that survive that correlation.

\subsubsection{Low energy range (30--200 eV)}

\begin{sidewaysfigure}
 \center{\includegraphics[width=8.5in]{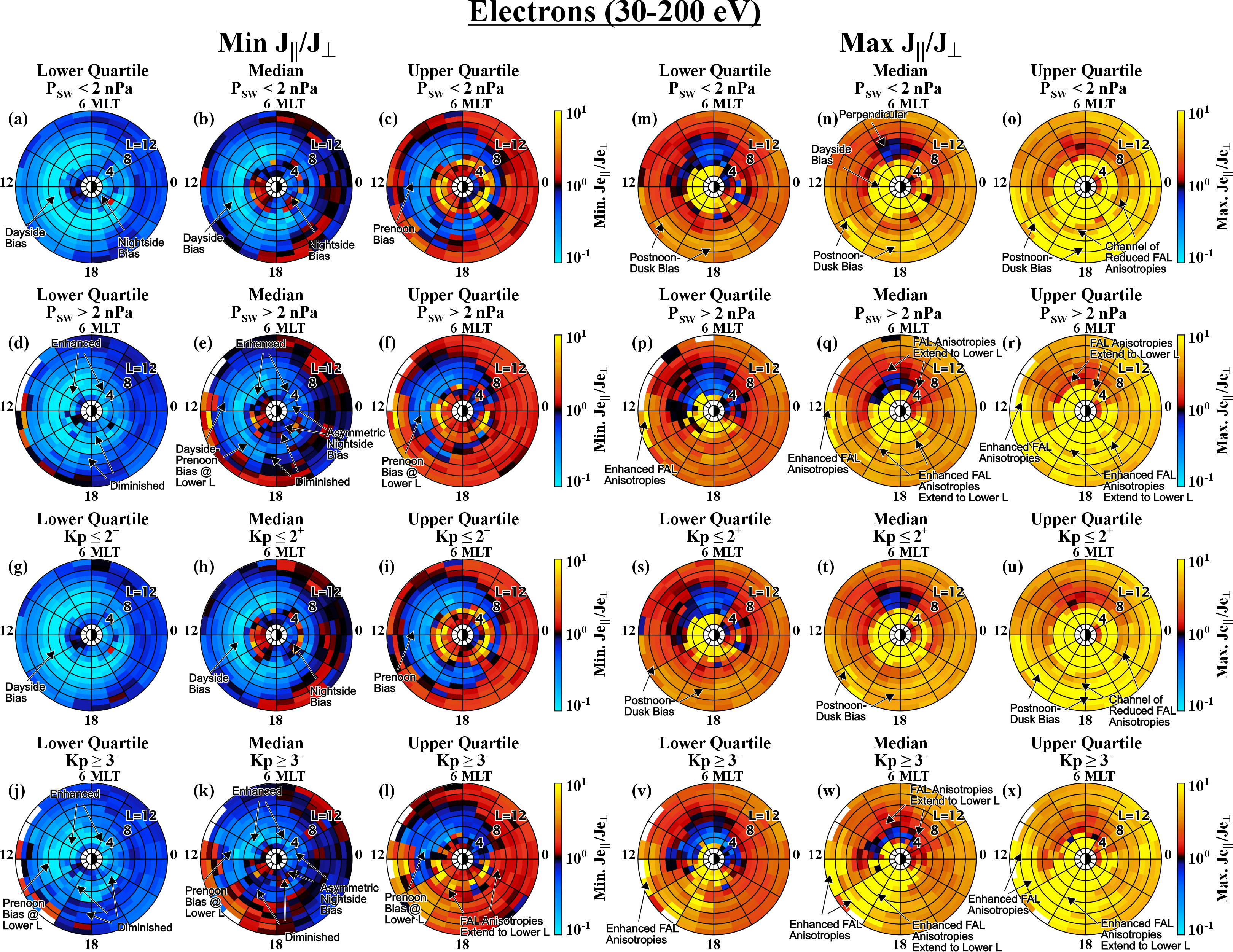}} 
\caption{The lower quartile, median, and upper quartile distributions of minimum electron flux ratios $J_{e\parallel}/J_{e\perp}$ in the energy range 30 $<$ E $<$ 200 eV for (a-c) $P_{\rm SW} \le 2$ nPa, (d-f) $P_{\rm SW} > 2$ nPa, (g-i) $Kp \le 2^+$, and (j-l) $Kp \ge 3^-$, the lower quartile, median, and upper quartile distributions of maximum $J_{e\parallel}/J_{e\perp}$ for (m-o) $P_{\rm SW} \le 2$ nPa, (p-r) $P_{\rm SW} > 2$ nPa, (s-u) $Kp \le 2^+$, and (v-x) $Kp \ge 3^-$.}
 \label{fig:minmaxanisotropy_30-200eV}
 \end{sidewaysfigure}

Figure \ref{fig:minmaxanisotropy_30-200eV} shows the \amin\ and \amax\ maps at 30--200 eV sorted by $P_{\rm SW}$ (panels a--f and m--r) and by $Kp$ (panels g--l and s--x).
Under high $P_{\rm SW}$, the regions of preferential transverse anisotropy are confined closer to Earth, owing in part to increased dayside compression, indicated by a gross inward shift of the largest transverse anisotropies (light blue to cyan).  
The confinement of nightside values to lower $L$ is attributed in part to enhanced sunward convection, arising from high $P_{\rm SW}$--which can enhance dayside and nightside reconnection \cite{boudouridisetal:2007,boudouridisetal:2008,boudouridisetal:2021,conneretal:2014}--or in relation to high $Kp$, which is moderately correlated with $P_{\rm SW}$.
The confinement is not symmetric about noon: the outer boundary from noon through dusk into the nightside has regressed further inward than on the prenoon and dawn sides, producing an asymmetric dayside/prenoon bias that is most apparent in the median and upper quartile maps (Figures \ref{fig:minmaxanisotropy_30-200eV}e and \ref{fig:minmaxanisotropy_30-200eV}f).
This asymmetry is associated with the expansion of isotropic (black) and FAL (red) regions to lower $L$ on the duskside and in the predawn sector accompanied by reductions of transverse anisotropies on the duskside and enhancements on the dawnside.
The plasmaspheric portion at $L\le$~4 in the lower and median quartile maps also acquires a morning/nightside bias in contrast to the symmetric nightside bias that occurs there under low $P_{\rm SW}$.
These effects point to enhanced FAL fluxes over the full band on the duskside and in the morning sector during high $P_{\rm SW}$, a contention borne out by the \amax\ maps.

Under high $P_{\rm SW}$, the \amax\ maps (Figures \ref{fig:minmaxanisotropy_30-200eV}p--\ref{fig:minmaxanisotropy_30-200eV}r) show enhanced parallel anisotropies at $L\gtrsim 10$ extending from dusk into the prenoon sector ($\sim$11--16 MLT), and enhancements in the dusk sector reaching lower $L$ than under low pressure.
This is most apparent in the upper quartile map, where the channel of reduced anisotropies occurring on the duskside at $L\sim$6--9 in Figure \ref{fig:minmaxanisotropy_30-200eV}o is effectively eliminated in Figure \ref{fig:minmaxanisotropy_30-200eV}r, producing a global plume-like extension of the FAL anisotropies.
FAL anisotropies outside the plasmasphere also extend to lower $L$ on the dawnside, but without appreciable enhancement in value.
This is accompanied by the regression of the plasmaspheric region to lower $L$ there.
The localized dawnside transverse region identified in Section \ref{sec:aniso_dist} also moves closer to Earth and spreads in MLT relative to its low pressure counterpart (compare Figures \ref{fig:minmaxanisotropy_30-200eV}p and \ref{fig:minmaxanisotropy_30-200eV}q with \ref{fig:minmaxanisotropy_30-200eV}m and \ref{fig:minmaxanisotropy_30-200eV}n).
These \amax\ trends map one-to-one onto the \amin\ trends described above: the duskside FAL enhancement coincides with the depressed transverse anisotropies of Figures \ref{fig:minmaxanisotropy_30-200eV}d and \ref{fig:minmaxanisotropy_30-200eV}e, while the unenhanced dawnside FAL region coincides with enhanced transverse anisotropies there.

Under high $Kp$, the same qualitative behavior occurs in the maps of \amin\ (Figures \ref{fig:minmaxanisotropy_30-200eV}j--\ref{fig:minmaxanisotropy_30-200eV}l) and \amax\ (\ref{fig:minmaxanisotropy_30-200eV}v--\ref{fig:minmaxanisotropy_30-200eV}x) but is systematically more pronounced.
The prenoon bias in the outer magnetosphere and the morning/nightside bias within the plasmasphere are both stronger in the \amin\ maps than under high $P_{\rm SW}$.
In the median map, a broader region of orange and red occurs from noon to dusk at $L\gtrsim$~8--10 than in the high $P_{\rm SW}$ case, and in the upper quartile map the FAL anisotropy floor levels span most $L$ on the duskside, with the largest floor levels occurring from postnoon to 16 MLT at $L \gtrsim 8$.
In addition to dayside compression, the nightside boundary of the \amin\ maps moves earthward under high $Kp$ owing to enhanced sunward convection in the equatorial plane.
The most distinctive high-$Kp$ feature appears in the median \amax\ map (Figure \ref{fig:minmaxanisotropy_30-200eV}w): a wedge-like channel of enhanced FAL anisotropy extending from the plasmasphere out to $L\sim$14 near 14--18~MLT, together with a strong enhancement just in the vicinity of noon at $L\ge 10$.
The wedge is absent under low $Kp$ and high $P_{\rm SW}$ (Figures \ref{fig:minmaxanisotropy_30-200eV}t and \ref{fig:minmaxanisotropy_30-200eV}q), and the noon enhancement is less pronounced under high $P_{\rm SW}$.
The high \amax\ values within the dawnside plasmasphere are also confined to lower $L$ under active conditions, with the boundary reaching $L\sim$3 under high $Kp$ against $L\sim$4 under high $P_{\rm SW}$.
This is accompanied by the inward intrusion and MLT spreading of the transverse regions just outside.
Also, unlike on the duskside, the dawnside FAL anisotropies outside the plasmasphere are not notably enhanced as their domain extends inward.

\subsubsection{Middle energy range (0.2--1 keV)}

 \begin{sidewaysfigure}
 \center{\includegraphics[width=8.5in]{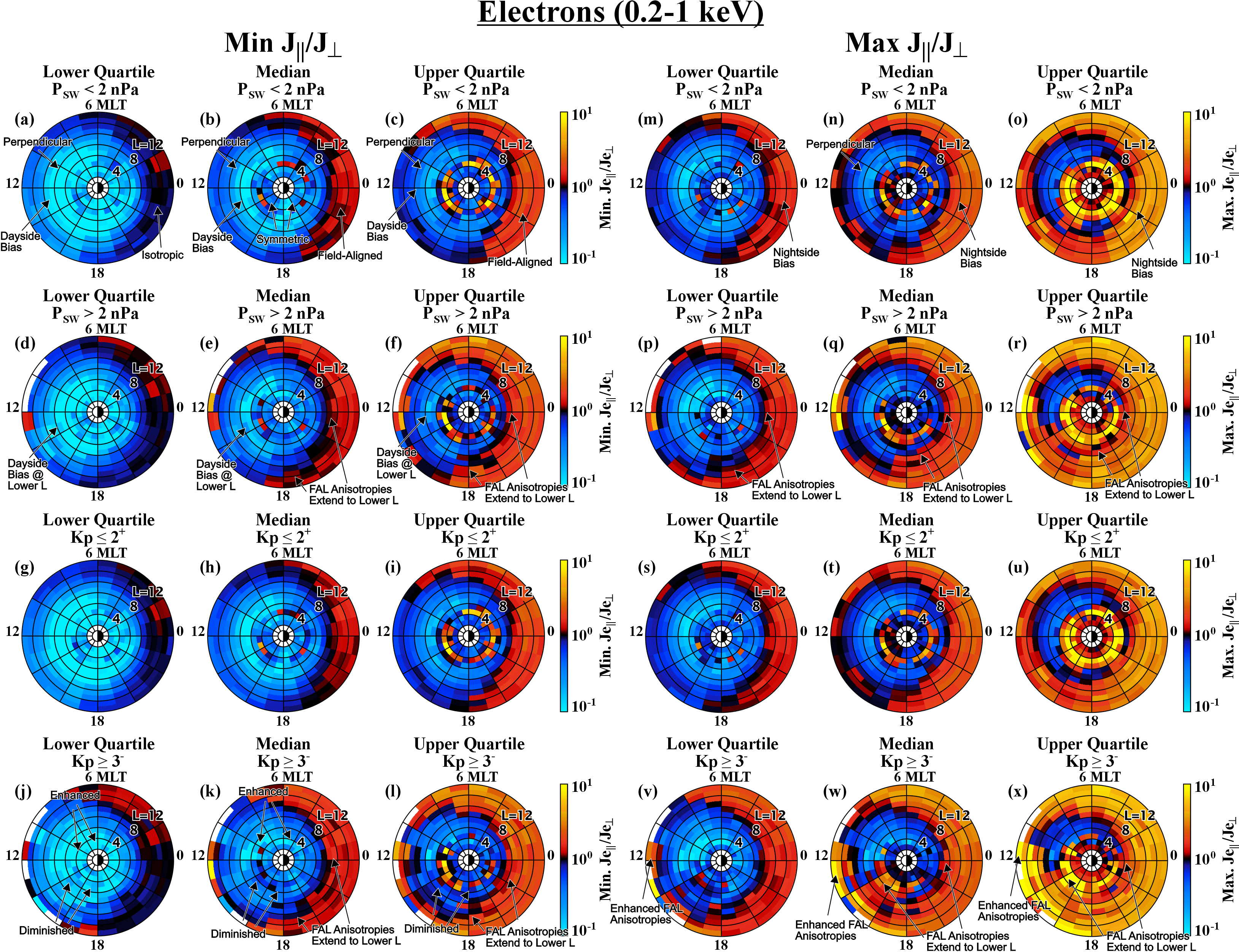}} 
\caption{The lower quartile, median, and upper quartile distributions of minimum electron flux ratios $J_{e\parallel}/J_{e\perp}$ in the energy range 0.2 $<$ E $<$ 1 keV for (a-c) $P_{\rm SW} \le 2$ nPa, (d-f) $P_{\rm SW} > 2$ nPa, (g-i) $Kp \le 2^+$, and (j-l) $Kp \ge 3^-$, the lower quartile, median, upper quartile distributions of maximum $J_{e\parallel}/J_{e\perp}$ for (m-o) $P_{\rm SW} \le 2$ nPa, (p-r) $P_{\rm SW} > 2$ nPa, (s-u) $Kp \le 2^+$, and (v-x) $Kp \ge 3^-$.}
 \label{fig:minmaxanisotropy_200-1000eV}
 \end{sidewaysfigure}

Figure \ref{fig:minmaxanisotropy_200-1000eV} shows the corresponding maps for 0.2--1.0 keV.
Under high $P_{\rm SW}$, the transverse anisotropy regions above $L>4$ retain their symmetric dayside bias but move closer to Earth at all MLT, an effect attributable to the combined action of increased compression--which shifts the regions of high transverse anisotropy (cyan) inward on the dayside--, sunward convection--which drive nightside anisotropies inward-, and enhanced FAL fluxes over the full band.
The latter is most apparent in the median and upper quartile maps (Figures \ref{fig:minmaxanisotropy_200-1000eV}e and \ref{fig:minmaxanisotropy_200-1000eV}f), which show a notable expansion of FAL anisotropy floor regions (orange to red) to lower $L$, predominantly on the nightside and the dawn and dusk flanks, and also further into the dayside.
A FAL floor region also appears near local noon that is absent under low $P_{\rm SW}$.
Correspondingly, the \amax\ maps under high $P_{\rm SW}$ (Figures \ref{fig:minmaxanisotropy_200-1000eV}p--\ref{fig:minmaxanisotropy_200-1000eV}r) show an expansion of the FAL domain to lower $L$ at most MLT with enhanced values, accompanied by the reduction of the transverse floor domain at all MLT that is most pronounced in the dusk sector, yielding a stronger prenoon bias.

Under high $Kp$, the transverse regions are likewise confined closer to Earth at all MLT, but here--unlike the high $P_{\rm SW}$ case--they develop a preferential prenoon bias, with diminished values from noon to dusk and enhanced values from dawn to noon at lower $L$.
This mirrors the low energy behavior but is less pronounced.
The FAL floor regions expand more broadly and to relatively more enhanced values than under high $P_{\rm SW}$.
The \amax\ maps under high $Kp$ (Figures \ref{fig:minmaxanisotropy_200-1000eV}v--\ref{fig:minmaxanisotropy_200-1000eV}x) show a marked increase in both FAL anisotropy amplitude and spatial domain, extending to lower $L$ and further into the dayside, and larger than under high $P_{\rm SW}$, being most pronounced from prenoon to dusk in the median and upper quartile maps.
As in the low energy range, the median map exhibits enhanced FAL anisotropies near local noon at $L \gtrsim 10$ and a wedge-like channel at 14--18 MLT extending down to the plasmaspheric region, albeit at reduced amplitude.
Again, the wedge is absent under low $Kp$ and high $P_{\rm SW}$, and the noon enhancement is less pronounced under high $P_{\rm SW}$.
The upper quartile map shows the same enhancements without an obvious wedge.

\subsubsection{High energy range (1--30 keV)}

 \begin{sidewaysfigure}
 \center{\includegraphics[width=8.5in]{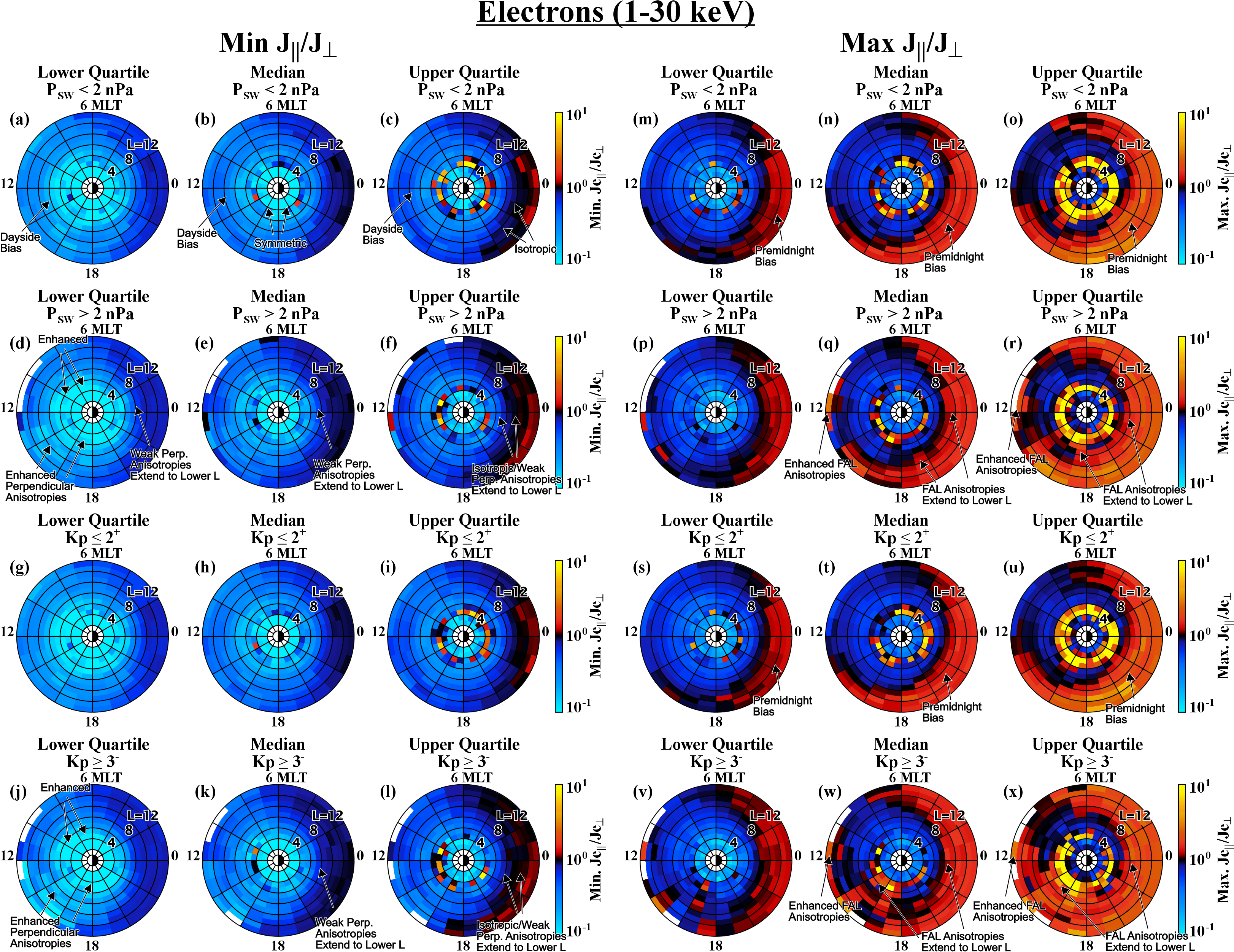}} 
\caption{The lower quartile, median, and upper quartile distributions of minimum electron flux ratios $J_{e\parallel}/J_{e\perp}$ in the energy range 1 $<$ E $<$ 30 keV for (a-c) $P_{\rm SW} \le 2$ nPa, (d-f) $P_{\rm SW} > 2$ nPa, (g-i) $Kp \le 2^+$, and (j-l) $Kp \ge 3^-$, the lower quartile, median, upper quartile distributions of maximum $J_{e\parallel}/J_{e\perp}$ for (m-o) $P_{\rm SW} \le 2$ nPa, (p-r) $P_{\rm SW} > 2$ nPa, (s-u) $Kp \le 2^+$, and (v-x) $Kp \ge 3^-$.}
 \label{fig:minmaxanisotropy_1-30keV}
\end{sidewaysfigure}

Figure \ref{fig:minmaxanisotropy_1-30keV} shows the 1--30 keV maps.
In all panels, the most strongly transverse regions (blue) occur from dawn to noon at $L\sim 6$--14 and at $L\sim 6$--8 at all other MLT, irrespective of condition, and the \amin\ maps retain a dayside bias under both low and high $P_{\rm SW}$.
Under high $P_{\rm SW}$, there is an inward shift of the transverse regions on both the dayside and nightside, indicated by reduced anisotropy values at $L > 10$ near local noon and by the inward shift of the isotropic and FAL regions on the nightside, together with enhanced transverse values at low $L$ on the dayside and higher values postnoon.
The yellow and red near $L\sim$4 is the radiation belt contamination artifact noted in Section \ref{sec:instrument}.
The high $Kp$ maps (Figures \ref{fig:minmaxanisotropy_1-30keV}j--\ref{fig:minmaxanisotropy_1-30keV}l) show the same dayside bias and a similar inward shift and enhancement.

The \amax\ maps under low $P_{\rm SW}$ and low $Kp$ reproduce the premidnight-centered horseshoe described in Section \ref{sec:aniso_dist}, extending from postnoon, premidnight, and dawn at $L\gtrsim$~12, $L\gtrsim$~8, and $L\gtrsim$~10--12, respectively, with the transverse anisotropy floors showing a prenoon bias that spans $L\lesssim$~14 prenoon but only $L\lesssim$~8 at premidnight.
Under high $P_{\rm SW}$, the FAL regions extend to lower $L$ and further into the dayside toward local noon--more pronounced on the duskside in the upper quartile map and in the morning sector in the median map.
An enhanced FAL region appears near noon at $L\gtrsim$~11--12 that is absent under low $P_{\rm SW}$. 
The transverse floor regions are correspondingly confined closer to Earth.
Under high $Kp$, the same expansion occurs but is again more dramatic, with remnants of the wedge-like channel at 14--16 MLT reaching down to the plasmaspheric region in the median map (Figure \ref{fig:minmaxanisotropy_1-30keV}w) and a more pronounced extension to lower $L$ in the dusk and prenoon sectors in the upper quartile map than under high $P_{\rm SW}$.

Taken across all three energy ranges, the \amax\ comparisons indicate that the energy extent of FAL electrons and, by inference, the domain of their source mechanism(s) are broader under high $Kp$ than under low $Kp$, being most effective in the dusk sector of the outer magnetosphere.
A similar but systematically weaker behavior occurs under high $P_{\rm SW}$, so that $Kp$ exerts the stronger influence in all three energy ranges.

\subsection{Electron Anisotropy Occurrences}
\label{sec:occurrences}

\begin{figure}
\center{\includegraphics[width=6.5in]{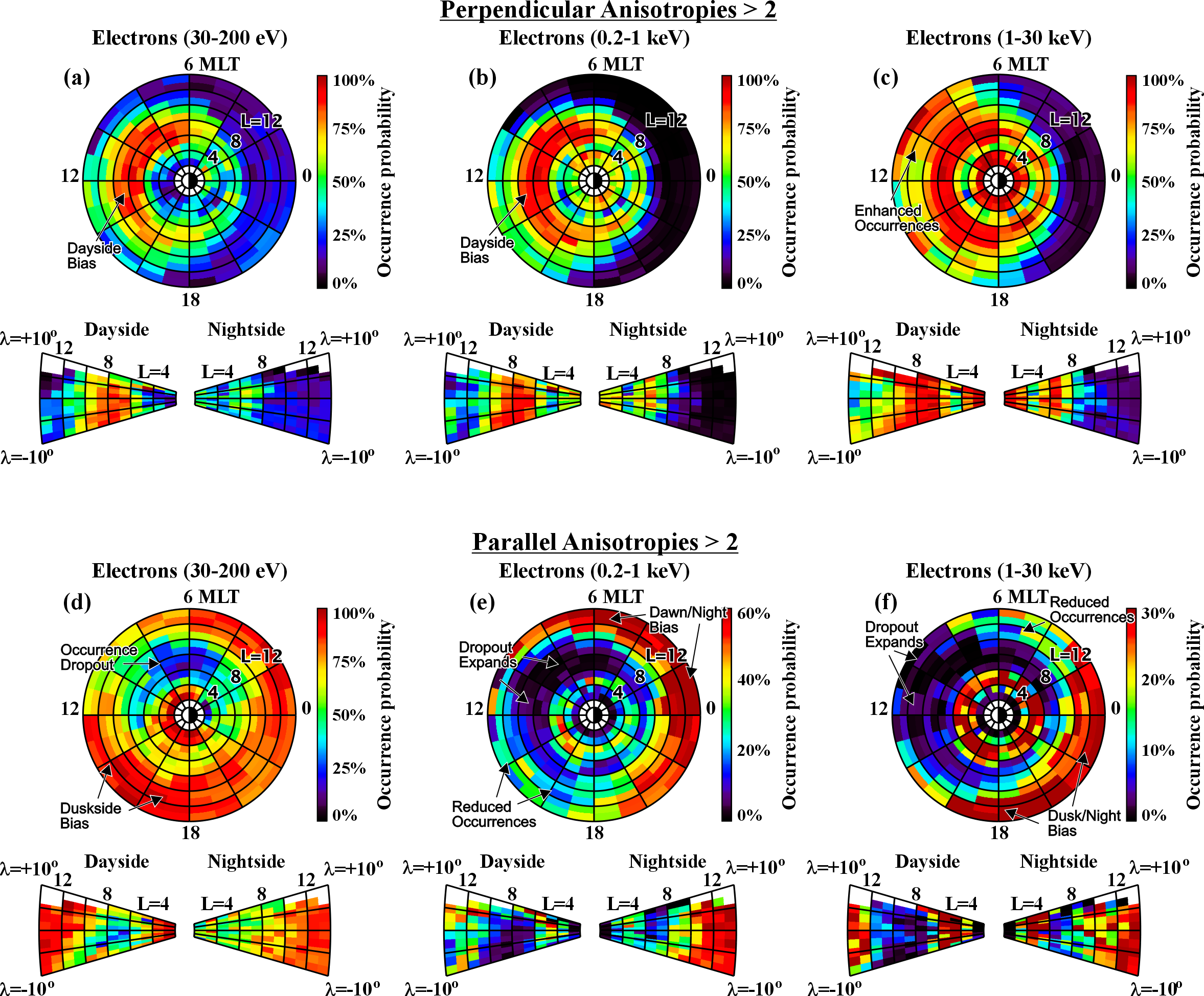}}
\caption{Occurrence probabilities of perpendicular electron anisotropies $>$ 2 (top row) as a function of $L$-MLT and (bottom row) $L$-$\lambda$ in the dayside and nightside for (a) 30-200 eV, (b) 0.2- 1 keV, and (c) 1-30 keV energy ranges; occurrence probabilities of parallel electron anisotropies $>$ 2 (top row) as a function of $L$-MLT and (bottom row) $L$-$\lambda$ in the dayside and nightside for (d) 30-200 eV, (e) 0.2- 1 keV, and (f) 1-30 keV energy ranges.}
\label{fig:occurrences}
\end{figure}

Figure \ref{fig:occurrences} shows occurrence probabilities of moderate to high anisotropies: panels a--c give the perpendicular case, defined per $L$-MLT (or $L$-$\lambda$) bin as the fraction of 3~s measurements with $\amin < 0.5$ (equivalently $\amin^{-1} > 2$), and panels d--f the parallel case, defined as the fraction with $\amax > 2$, in each case expressed as a percentage.
The top panel of each set gives the distribution in $L$-MLT and the lower two give dayside and nightside distributions in $L$-$\lambda$.
Because of the low inclination orbits of THEMIS A and D, the $L$-$\lambda$ distributions are limited to $|\lambda|\le$10\textdegree. 
They are nevertheless informative about dayside-nightside differences and latitudinal trends.

\subsubsection{Preferential perpendicular anisotropy case}

Figures \ref{fig:occurrences}a--\ref{fig:occurrences}c show a clear dayside bias at $L\gtrsim 6$, with probabilities reaching $\ge$~75\% in each energy range, attributable to the compressed dayside/stretched nightside field configuration.
Both the probability values and their $L$ extent increase systematically with energy.
In the low energy range, the largest dayside probabilities occupy a single broad band at $L\sim$6--10, reducing beyond $L>$12 except near local noon.
Probabilities are significantly depressed on the dayside at $L\le 4$.
In the high energy range, by contrast, probabilities at $L\le 4$ are high (median 94\%) and are also elevated at $L\ge 5$ (median 92\% at $L\sim$6--9).
At $L\ge 9$ the high energy distribution becomes asymmetric, with significantly higher values prenoon ($\ge$~75\%) than postnoon ($\sim$50--60\%).
The corresponding $L$-$\lambda$ panel shows that this MLT asymmetry is accompanied by an asymmetry about $\lambda=0$\textdegree, the highest values occurring near $|\lambda|\sim$~10\textdegree, suggesting a latitudinal dependence for this class of transverse electrons.
The mid energy range is intermediate throughout.

On the nightside, the same trend of increasing probability with energy occurs, but with moderate to high values confined to lower $L$ than on the dayside.
The low energy range shows two bands of moderate probability, at $L\le$~4 and $L\sim$5--8; the former is elevated relative to its dayside counterpart, giving the night/morning bias already evident in the low energy \amin\ maps.
In the high energy range, both bands reach $\ge$~70\% (median 83\%), and the $L\le$4 band becomes roughly symmetric with its dayside counterpart.
Unlike on the dayside, no preferential enhancement at higher $|\lambda|$ occurs.
At $L\gtrsim 8$ on the nightside, probabilities are depressed ($\lesssim$~25\%) in all three energy ranges, reflecting the more isotropic and/or field-aligned distributions there (Figures \ref{fig:minmaxanisotropy_lo-high_e}b, \ref{fig:minmaxanisotropy_lo-high_e}e, and \ref{fig:minmaxanisotropy_lo-high_e}h).
The narrow band of depressed probabilities near $L\sim$~4--5 is affected by the radiation belt contamination noted in Section \ref{sec:instrument} and is not interpreted, although it does provide a useful visual delimiter between the plasmasphere and the outer magnetosphere.

\subsubsection{Preferential parallel anisotropy case}

In contrast to the perpendicular case, Figures \ref{fig:occurrences}d--\ref{fig:occurrences}f show that the probability of moderate to high parallel anisotropies \emph{decreases} with increasing energy (note the scale change with energy), indicating that FAL anisotropies are dominated by low energy FAL electrons with non-negligible contributions from mid- and high-energy electrons.
The three patterns also differ substantially from one another, indicative of changing FAL electron populations and/or energization mechanisms.

In the low energy range (Figure \ref{fig:occurrences}d) the $L$-MLT pattern exhibits a duskside bias in the outer magnetosphere, with $\gtrsim$~75\% probabilities extending over $L\sim 8$--14 against $L\sim 10$--14 on the dawnside, forming an asymmetric horseshoe centered on premidnight.
The broader duskside region reflects the highly field-aligned electrons with $\amax\sim$5--10 seen in Figures \ref{fig:minmaxanisotropy_lo-high_e}j and \ref{fig:minmaxanisotropy_lo-high_e}k.
Moderately elevated probabilities (50--60\%) extend from $L\sim 10$ down to $L\sim$3--4 in the morning sector and prenoon at $L\sim 8$--14.
A pronounced drop ($\lesssim$~30\%) appears at $L\sim$5--10 on the dawnside, coinciding exactly with the localized region where transverse anisotropies extend over the full low energy band in Figures \ref{fig:minmaxanisotropy_lo-high_e}j and \ref{fig:minmaxanisotropy_lo-high_e}k, and confirming the dawnside void identified in Section \ref{sec:aniso_dist}.
Enhanced occurrences also appear at $L \lesssim 4$ with a dayside/postnoon bias, corresponding to the high plasmaspheric FAL anisotropies of Figures \ref{fig:minmaxanisotropy_lo-high_e}j--\ref{fig:minmaxanisotropy_lo-high_e}l.
In $L$-$\lambda$ space (Figure \ref{fig:occurrences}d, lower panels), the high-probability region at $L\le$~4 is roughly symmetric about $\lambda = 0$, whereas at $L \ge$~8--12 the dayside distribution is strongly asymmetric with peak values trending toward positive $\lambda$ and the nightside distribution shows a slight opposite trend.

In the mid energy range (Figure \ref{fig:occurrences}e), the horseshoe becomes asymmetric about the morning sector at $L\ge 8$, with the premidnight-to-midnight region spanning $L\sim 8$--14 against $L\sim 10$--14 in the morning and dawn sectors.
Probabilities within the horseshoe ($\sim$60\%) are reduced relative to the low energy range but remain significant, indicating that sizable FAL anisotropies often extend from the low into the mid energy range in these regions.
Values in the noon-dusk sector at $L\ge 8$ are diminished (20--40\%) relative to the low energy range, so that the pattern acquires a dawnside/nightside bias.
The dawnside dropout expands substantially in both MLT and $L$, with probabilities falling to $\lesssim$~20\%, consistent with the expansion of the region dominated by transverse anisotropies over the full mid energy band (Figures \ref{fig:minmaxanisotropy_lo-high_e}m--\ref{fig:minmaxanisotropy_lo-high_e}o).
Probabilities also drop at $L\lesssim$4, where the low energy range showed enhancements, because the mid energy FAL anisotropies there fall below the threshold of 2 (Figure \ref{fig:minmaxanisotropy_lo-high_e}o).
The $L$-$\lambda$ panels show that the diminished dayside probabilities from noon to dusk at $L>8$ cluster near $\lambda = 0$\textdegree, whereas the enhanced values from dawn to noon ($\sim$60\%) are offset to $|\lambda|\sim$5--10\textdegree.
The nightside counterparts are roughly symmetric about $\lambda=$0\textdegree.

In the high energy range (Figure \ref{fig:occurrences}f) the horseshoe is centered on premidnight at $L\ge 8$, and the probabilities are further reduced, with the highest values limited to $\sim$25--30\% in the dusk, midnight, and morning sectors at $L\gtrsim 10$.
Opposite to the mid energy range, lower but non-negligible levels ($\sim 15\%$) occur on the dawnside at $L\gtrsim 10$, giving a dusk/night bias.
The dropout region has expanded to larger $L$ prenoon, with probabilities falling to $\lesssim$~10\%, and a further dropout occurs at $L\lesssim 4$, where FAL anisotropies are largely absent in this range (Figures \ref{fig:minmaxanisotropy_lo-high_e}p--\ref{fig:minmaxanisotropy_lo-high_e}r).
The ring of elevated values near $L=4$ is the radiation belt artifact.
In $L$-$\lambda$ space, the highest values at $L>$8 are roughly symmetric about $\lambda = 0$\textdegree, and the dayside dropout is more pronounced, spanning $L = 6$--12.

The systematic change of pattern with energy in Figures \ref{fig:occurrences}d--f indicates changes in the contributing FAL electron populations.
Comparing the three, FAL electrons in the late-morning, dusk, and prenoon sectors primarily span 30~eV--1~keV; those in the postnoon to dusk sectors are primarily limited to the low energy range with non-negligible mid and high energy contributions; and those from premidnight to early morning are dominated by the low energy range with important mid and high energy contributions.
Finally, comparison of Figures \ref{fig:occurrences}d--f with Figures \ref{fig:occurrences}a--c shows that the FAL and transverse populations are close to mutually exclusive in each of the three energy ranges.

\subsection{Occurrence Dependencies on Solar Wind Dynamic Pressure and $Kp$}
\label{sec:occ_psw_kp}

\subsubsection{Preferential perpendicular anisotropy case}

\begin{figure}
\center{\includegraphics[width=6.5in]{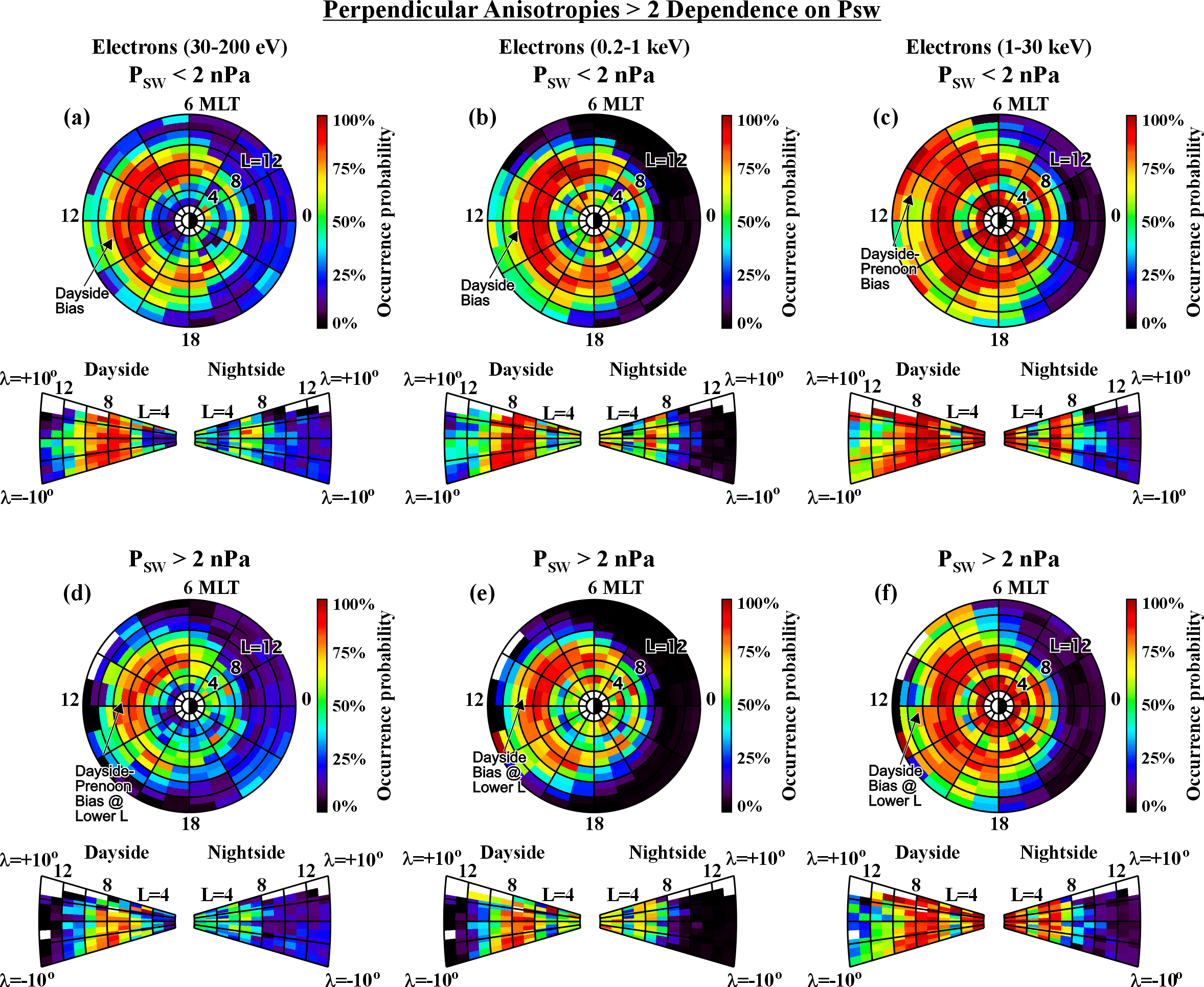}} %
\caption{Occurrence probabilities of perpendicular electron anisotropies $>$ 2 (minimum flux ratios $J_{e\parallel}/J_{e\perp} < 0.5$) as a function of (top panel) $L$-MLT and (bottom two panels) $L$-$\lambda$ in the dayside and nightside for (a-c) $P_{\rm SW} \le 2$ nPa and (d-f) $P_{\rm SW} > 2$ nPa in the (a, d) 30-200 eV, (b, e) 0.2- 1 keV, and (c,f) 1-30 keV energy ranges.
}
\label{fig:anisper_occurrences_psw}
\end{figure}

\begin{figure}
\center{\includegraphics[width=6.5in]{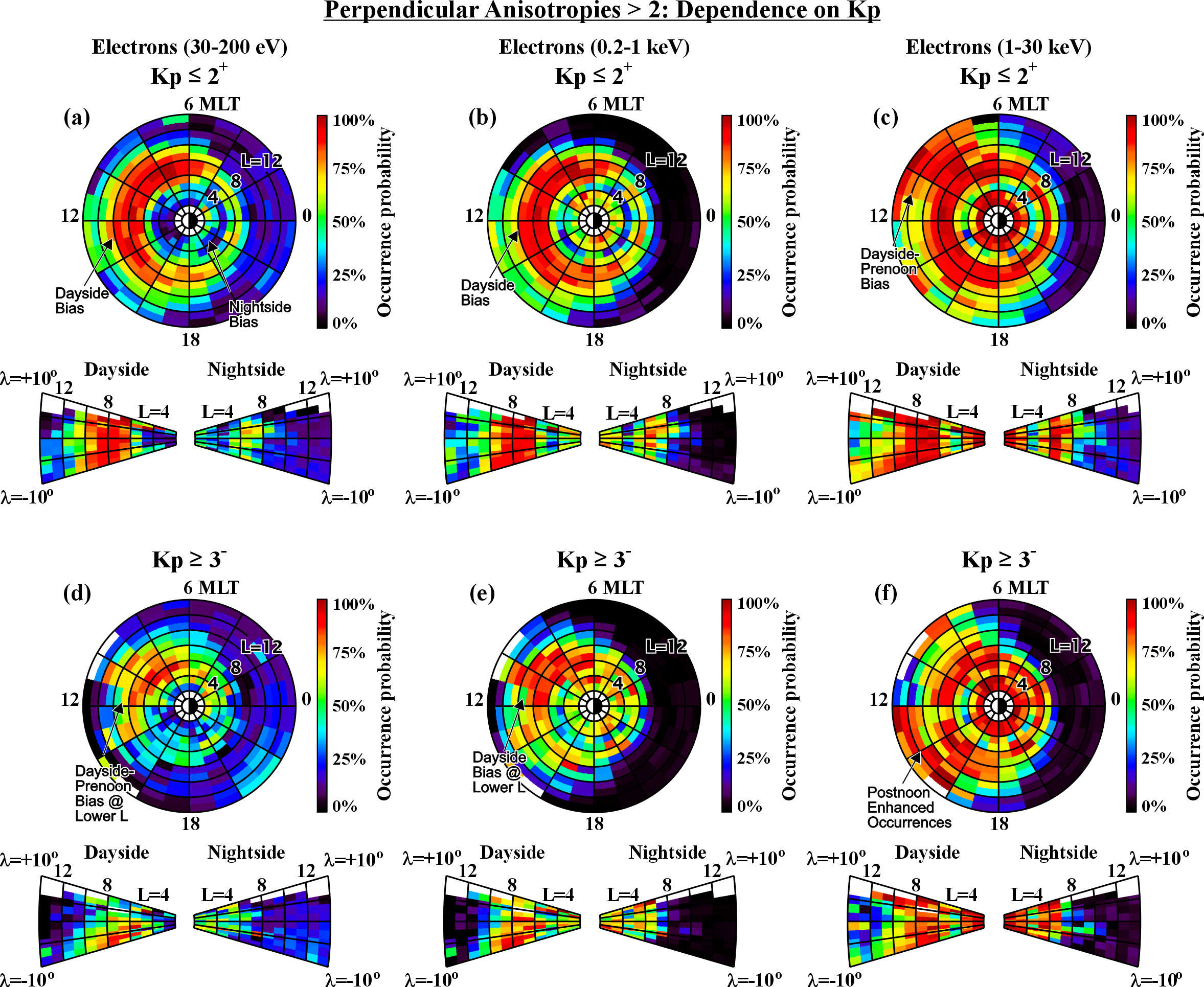}} %
\caption{Occurrence probabilities of perpendicular electron anisotropies $>$ 2 (minimum flux ratios $J_{e\parallel}/J_{e\perp} < 0.5$) as a function of (top panel) $L$-MLT and (bottom two panels) $L$-$\lambda$ in the dayside and nightside for (a-c) $Kp \le 2^+$, and (d-f) $Kp \ge 3^-$ in the (a, d) 30-200 eV, (b, e) 0.2- 1 keV, and (c,f) 1-30 keV energy ranges.
}
\label{fig:anisper_occurrences_kp}
\end{figure}

Figures \ref{fig:anisper_occurrences_psw} and \ref{fig:anisper_occurrences_kp} show the transverse anisotropy occurrence probabilities sorted by $P_{\rm SW}$ and by $Kp$, respectively, in the same format as Figure \ref{fig:occurrences}.
Consistent with Section \ref{sec:aniso_psw_kp}, the low $P_{\rm SW}$ and low $Kp$ distributions (Figures \ref{fig:anisper_occurrences_psw}a--\ref{fig:anisper_occurrences_psw}c and \ref{fig:anisper_occurrences_kp}a--\ref{fig:anisper_occurrences_kp}c) closely resemble the undifferentiated distributions of Figures \ref{fig:occurrences}a--\ref{fig:occurrences}c: elevated dayside values ($\ge$~75\%) spanning $L\sim$5--12, moderate to high nightside values at $L\sim$6--8, the same increase of probability and $L$ extent with energy, and the same transition at $L\lesssim 4$ from moderate, nightside-biased probabilities at low energy to high, MLT-symmetric probabilities at high energy.

Under high $P_{\rm SW}$, the dayside bias persists but the enhanced region ($\ge$~50\%) is confined to $L\sim$5--11 or 12 against $L\sim$5--14 under low pressure, consistent with increased magnetospheric compression, and the highest values ($\ge$~75\%) in the low and, to a lesser extent, mid energy ranges become more concentrated prenoon than postnoon.
In the high energy range, occurrence probabilities in the prenoon sector are significantly reduced (Figure \ref{fig:anisper_occurrences_psw}f).
In the nightside, regions of moderate to high occurrence probabilities have shifted inward by $\sim1 \Delta L$, attributed to enhanced sunward convection (since $P_{\rm SW}$ correlated with $Kp$) and the expansion of regions with only FAL anisotropies over the entire band.

Under high $Kp$, the same inward shift in the dayside and nightsie occurs in the low and mid energy ranges (Figures \ref{fig:anisper_occurrences_kp}d and \ref{fig:anisper_occurrences_kp}e), with the largest probabilities ($\gtrsim$~70\% and $\gtrsim$75\%, respectively) localized prenoon--more sharply localized than in the high $P_{\rm SW}$ case and asymmetric about $\lambda=$~0\textdegree\ extending to larger negative values with increasing $L$, while they are more symmetric under high $P_{\rm SW}$ case.
In the high energy range (Figure \ref{fig:anisper_occurrences_kp}f) prenoon values are significantly reduced relative to quiet conditions while postnoon values at $L\ge 10$ increase significantly, and the high-probability regions on the dayside ($L\le 10$) and nightside ($L\le 6$) lie closer to Earth (by $\sim1 \Delta L$) than under low $Kp$.

\subsubsection{Preferential parallel anisotropy case}

\begin{figure}
\center{\includegraphics[width=6.5in]{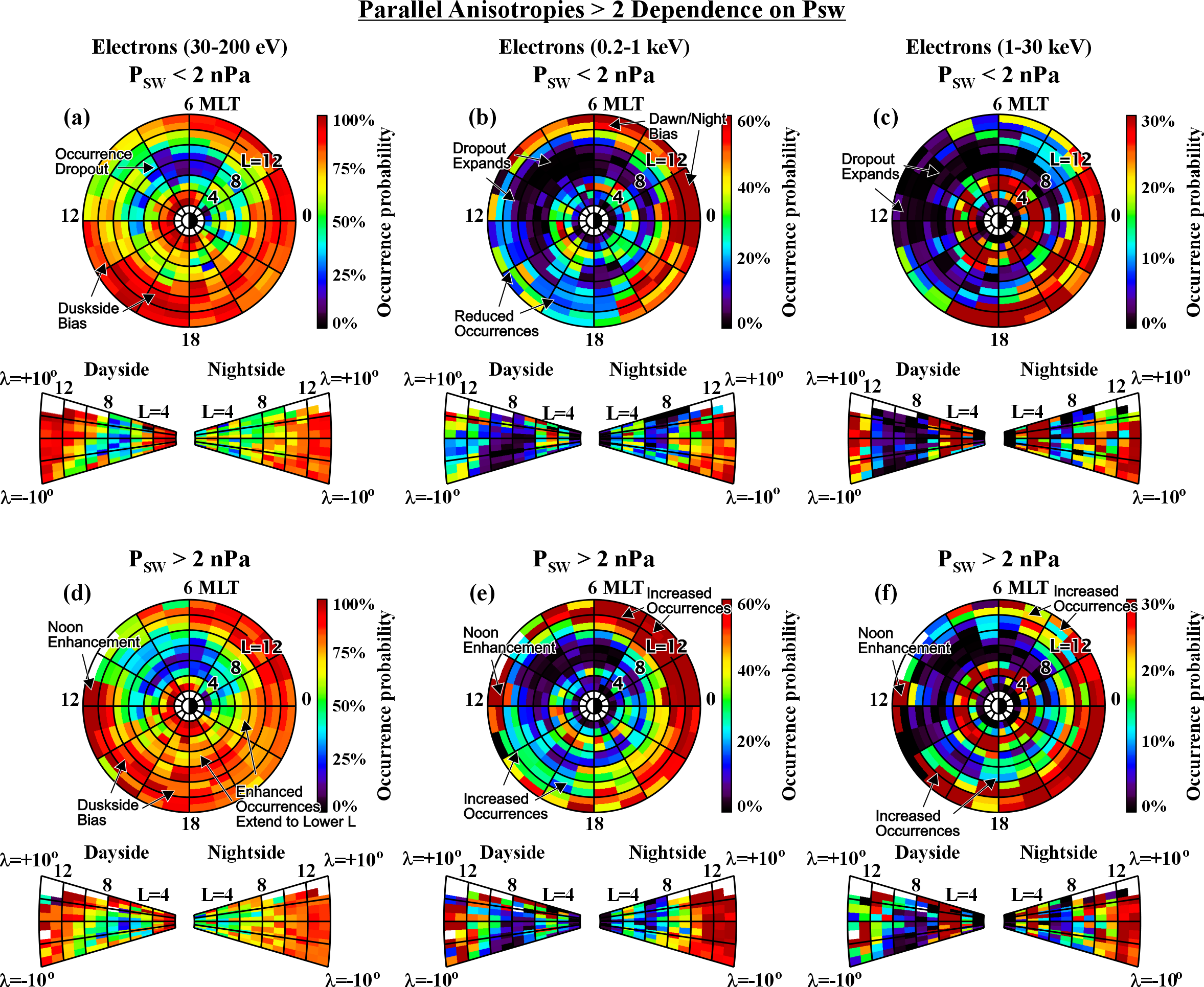}}
\caption{Occurrence probabilities of parallel electron anisotropies $>$ 2 (maximum flux ratios $J_{e\parallel}/J_{e\perp} > 2$) as a function of (top panel) $L$ and MLT and (bottom two panels) $L$ and $\lambda$ in the dayside and nightside for (a-c) $P_{\rm SW} \le 2$ nPa and (d-f) $P_{\rm SW} > 2$ nPa in the (a, d) 30-200 eV, (b, e) 0.2- 1 keV, and (c,f) 1-30 keV energy ranges.}
\label{fig:anispar_occurrences_psw}
\end{figure}

\begin{figure}
\center{\includegraphics[width=6.5in]{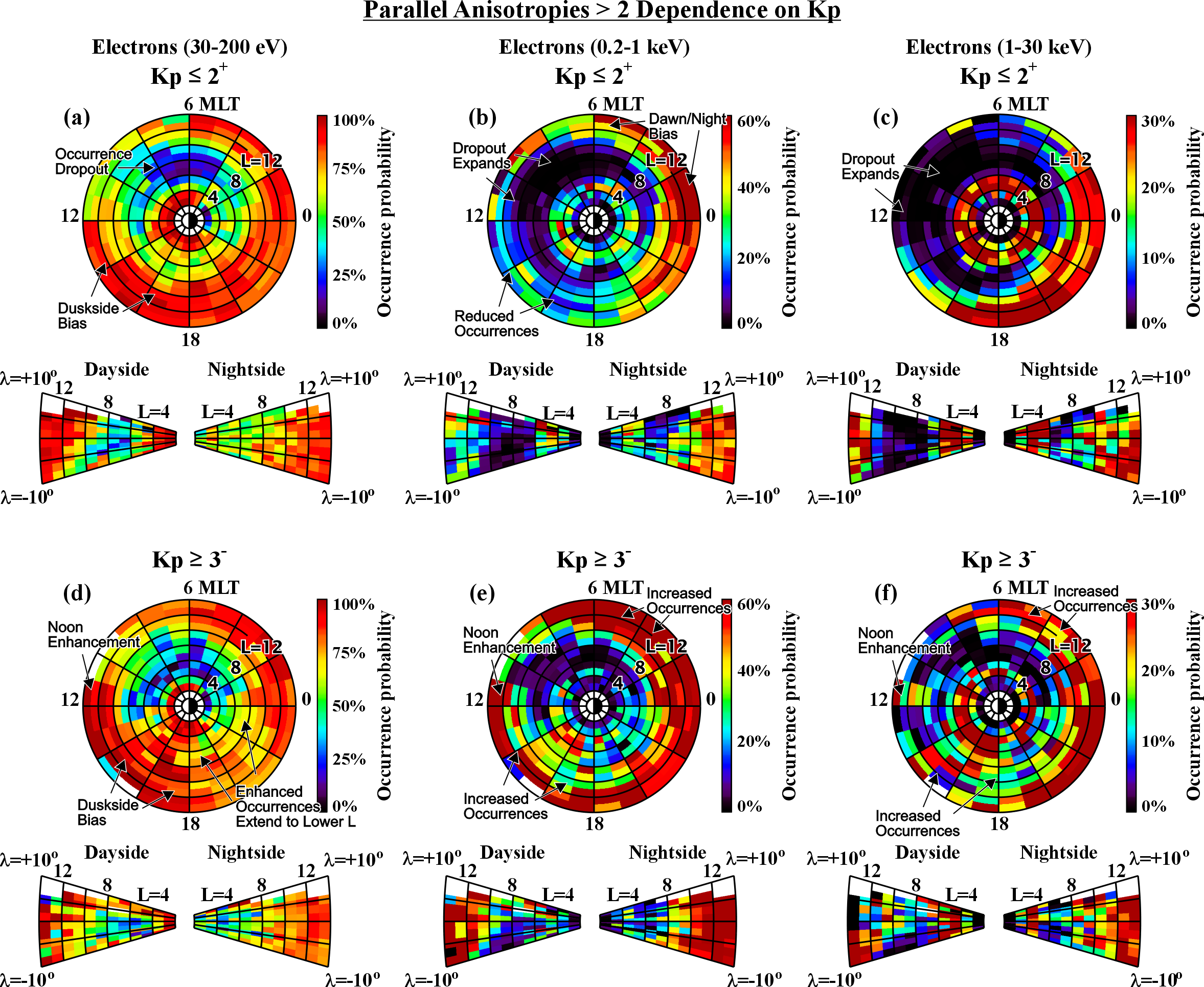}}
\caption{Occurrence probabilities of parallel electron anisotropies $>$ 2 (maximum flux ratios $J_{e\parallel}/J_{e\perp} > 2$) as a function of (top panel) $L$ and MLT and (bottom two panels) $L$ and $\lambda$ in the dayside and nightside for (a-c) $Kp \le 2^+$ and (d-f) $Kp \ge 3^-$ in the (a, d) 30-200 eV, (b, e) 0.2- 1 keV, and (c,f) 1-30 keV energy ranges.}
\label{fig:anispar_occurrences_kp}
\end{figure}

Figures \ref{fig:anispar_occurrences_psw} and \ref{fig:anispar_occurrences_kp} show the corresponding FAL anisotropy occurrence probabilities sorted by $P_{\rm SW}$ and $Kp$, in the same format as Figure \ref{fig:occurrences}.
As above, the low $P_{\rm SW}$ and low $Kp$ distributions closely resemble the undifferentiated ones of Figures \ref{fig:occurrences}d--\ref{fig:occurrences}f, reproducing the same outer magnetosphere biases (duskside in the low and high energy ranges, dawn/nightside in the mid energy range), the same dayside bias at $L\le$~4 in the low energy range that disappear in the other two ranges, the same decrease of probability with increasing energy, and the same dawnside dropout that expands with energy.

Under high $P_{\rm SW}$ (Figures \ref{fig:anispar_occurrences_psw}d--\ref{fig:anispar_occurrences_psw}f), probabilities are enhanced and the domain of high values ($\gtrsim$~70\%) extends to lower $L$--down to $L\sim$~4 at dusk, $L\sim$~6 at midnight, and $L\sim$~10 at dawn in the low energy range--together with a notable enhancement about local noon at $L\gtrsim$~10.
Enhancements over a large MLT extent, reaching lower $L$ and with a strong enhancement near local noon, also occur in the mid and high energy ranges.

Under high $Kp$ (Figures \ref{fig:anispar_occurrences_kp}d--\ref{fig:anispar_occurrences_kp}f) the same behavior occurs but is stronger.
In the low energy range the enhancements are most pronounced in the dusk sector, where high probabilities ($\gtrsim$~70\%) are seen down to $L\sim$~4, yielding a stronger duskside bias than under quiet conditions and higher values than in the high $P_{\rm SW}$ counterpart (Figure \ref{fig:anispar_occurrences_psw}d).
The strong enhancement about local noon at $L\gtrsim$~10 is also present.
Together with Section \ref{sec:aniso_psw_kp}, these comparisons establish that both $P_{\rm SW}$ and $Kp$ control the spatial domains of both anisotropy classes, and that $Kp$ exerts the stronger influence.

\section{Discussion}
\label{sec:discussion}

\begin{figure}
\center{\includegraphics[width=5.5in]{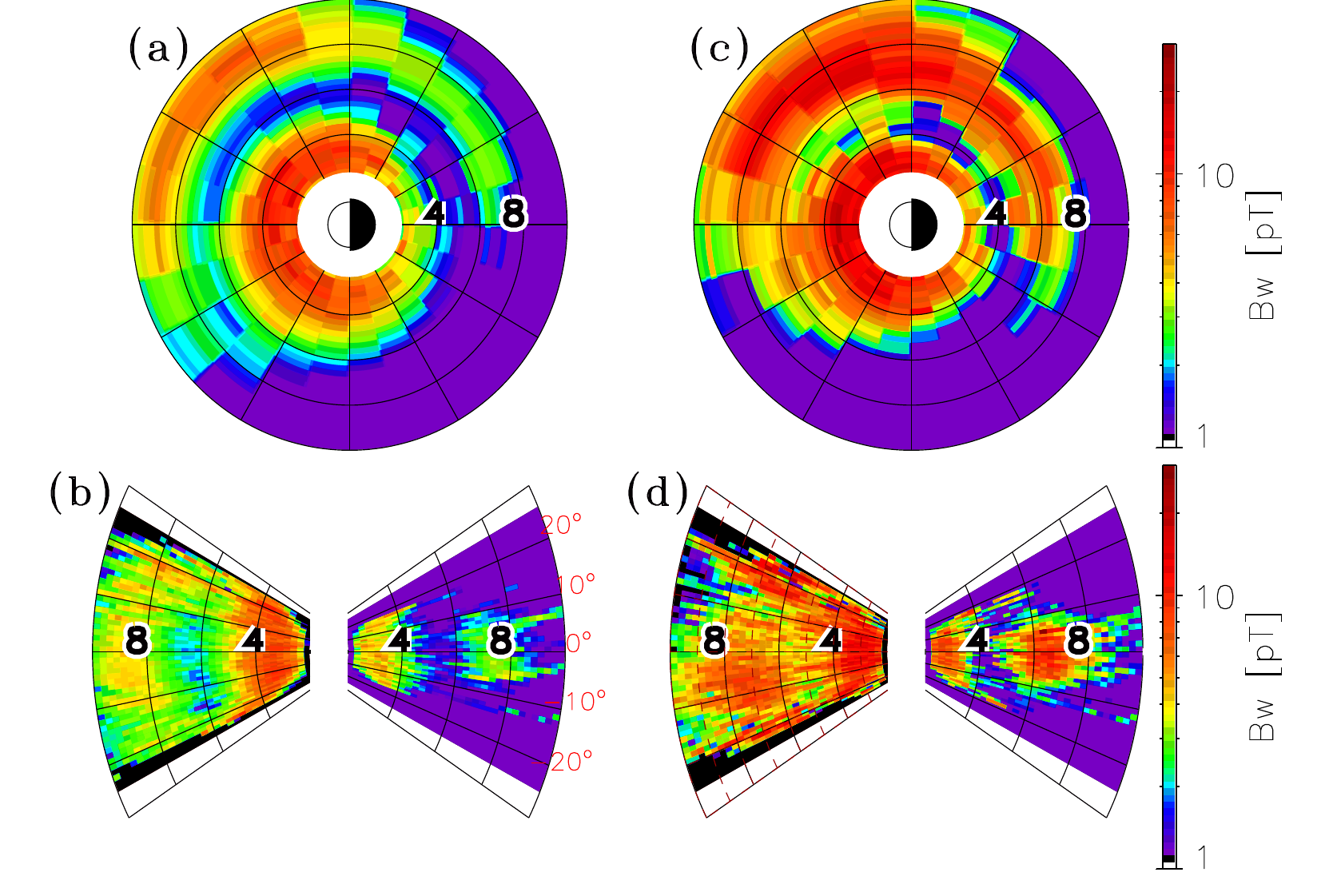}}
\caption{The distribution of lower band chorus wave magnetic field amplitude ($0.1f_{ce} < f < 0.5f_{ce}$) and hiss magnetic field amplitudes in (a,c) MLT and $L$ space and in $L$ and $\lambda$ on the (b,d) dayside and nightside for (a,b) low ($Kp < 3$) and (c,d) active ($Kp > 3$) magnetospheric conditions. Adapted from \citeA{agapitovetal:2018}.}
\label{fig:waves}
\end{figure}

The case study and statistical results presented in Sections \ref{sec:event} and \ref{sec:stats} demonstrate that the equatorial magnetosphere at $L\le14$ is populated by electron distributions with clearly distinguishable field-aligned and transverse pitch-angle (flux-ratio) anisotropy characteristics, whose spatial occurrences depend systematically on energy, MLT, $L$-shell, magnetic latitude, solar wind dynamic pressure, and geomagnetic activity level.
These anisotropies provide sources of free energy for whistler-mode chorus and hiss, and for electrostatic Time Domain Structures (TDS), and thus provide a useful observational basis for interpreting the global whistler wave distributions shown for comparison in Figure \ref{fig:waves}.

\subsection{Perpendicular Anisotropies and Quasi-Parallel Chorus}
The statistical distributions of \amin\ in Figure \ref{fig:minmaxanisotropy_lo-high_e} and the associated occurrence probabilities in Figures \ref{fig:occurrences}a--\ref{fig:occurrences}c show that preferential transverse anisotropies of electrons at 1--30~keV are concentrated in a narrow $L$ range on the nightside ($5<L<6$) and are more broadly distributed on the dayside, with the largest values and highest occurrence rates ($\ge$75\%) confined to $L\sim$5--10 (Figures \ref{fig:minmaxanisotropy_lo-high_e}g, \ref{fig:minmaxanisotropy_lo-high_e}h, and \ref{fig:minmaxanisotropy_lo-high_e}i).
This spatial pattern agrees with the distribution of chorus waves in Figure \ref{fig:waves} and is consistent with the picture in which the transverse anisotropy of energetic electrons ($T_\perp>T_\parallel$), continually resupplied by substorm injections from the magnetotail \cite{Birn2012,Chaston2014,Ergun2015}, drives cyclotron-resonant growth of quasi-parallel whistler-mode chorus \cite{kennel+petschek:1966,helliwell:1967,gary+wang:1996,Omura2009}.
We note that the population usually invoked for quasi-parallel chorus growth extends to $\sim$100~keV, whereas the present survey is limited to $\le$30~keV by the ESA energy range; the correspondence drawn here therefore applies to the low-energy portion of that population, and its extension to higher energies is an inference rather than a measurement.
Our finding that the transverse anisotropy occurrence probabilities and their spatial extent increase with electron energy across the measured range (compare Figures \ref{fig:occurrences}a, \ref{fig:occurrences}b, and \ref{fig:occurrences}c) is in the same sense as the reported dependence of dayside quasi-parallel chorus generation on tens-of-keV electrons \cite{Li2012,Meredith2012}, and the pronounced dayside/prenoon occurrence bias evident in Figures \ref{fig:occurrences}c and \ref{fig:anisper_occurrences_psw} is qualitatively consistent with the latitudinal broadening ($|\lambda|\sim40^\circ$) and off-equatorial peaking ($|\lambda|\sim5$--$10^\circ$) reported for dayside chorus that has evolved via azimuthal drift from the nightside injection region \cite{Agapitov2013,agapitovetal:2018,Li2016}.

The dependence of \amin\ and its occurrence probability on $P_{\rm SW}$ and $Kp$ (Figures \ref{fig:minmaxanisotropy_30-200eV}--\ref{fig:minmaxanisotropy_1-30keV} and \ref{fig:anisper_occurrences_psw}--\ref{fig:anisper_occurrences_kp}) offers a natural explanation for the sensitivity of dayside chorus wave amplitude and spatial extent to magnetospheric activity illustrated in Figures \ref{fig:waves}c and \ref{fig:waves}d.
The inward, prenoon-biased confinement of the largest transverse anisotropies under high $P_{\rm SW}$ and high $Kp$ (Figures \ref{fig:minmaxanisotropy_30-200eV}e,f and \ref{fig:anisper_occurrences_kp}d--f) parallels the compression of the dayside chorus source region to lower $L$ during active conditions seen in Figures \ref{fig:waves}c,d relative to quiet conditions in Figures \ref{fig:waves}a,b.
This supports the interpretation that magnetospheric compression, together with the enhanced FAL electron fluxes discussed below, jointly regulate both the anisotropy and the resulting quasi-parallel wave distributions.

The mid-energy (0.2--1~keV) transverse anisotropies documented in Section \ref{sec:stats} fall within the 0.1--10~keV range widely identified as the principal free-energy source for electron cyclotron harmonic (ECH) waves \cite{Horne2003}.
ECH waves, generated near the geomagnetic equator by the same anisotropy instability invoked above for quasi-parallel chorus, are considered a dominant driver of diffuse auroral electron precipitation through efficient pitch-angle scattering of 0.1--10~keV electrons into the loss cone \cite{Ni2016,Ni2011b}, complementing the role played by whistler-mode chorus at somewhat higher energies.
The dayside-biased occurrence pattern of mid-energy transverse anisotropies reported here, together with its confinement to $L\sim$5--14 and its intensification under active ($Kp\ge3^-$) conditions (Figures \ref{fig:occurrences}b and \ref{fig:anisper_occurrences_kp}b,e), is broadly consistent with statistical ECH wave occurrence and amplitude distributions, which likewise peak on the dayside, span a comparable range of $L$, and strengthen with geomagnetic activity \cite{Ni2016}.
The mid-energy anisotropy maps presented here may therefore serve as a useful observational proxy for regions favorable to ECH wave growth and, by extension, for diffuse auroral scattering.

\subsection{Field-Aligned Anisotropies, Oblique Chorus, and TDS}
In contrast to the transverse anisotropies, the FAL anisotropies diagnosed via \amax\ are largest and most probable at low energies (30--200~eV; Figures \ref{fig:minmaxanisotropy_lo-high_e}j--\ref{fig:minmaxanisotropy_lo-high_e}l and \ref{fig:occurrences}d), consistent with the case-study distributions in Figures \ref{fig:event1}f and \ref{fig:edist_event1}, which showed counterstreaming field-aligned electrons dominating the density at energies just above the spacecraft potential.
The decrease in occurrence probability with increasing energy from Figure \ref{fig:occurrences}d to Figure \ref{fig:occurrences}f indicates that FAL anisotropies are primarily carried by this low-energy population, with only limited overlap into the 0.2--1~keV and 1--30~keV ranges.
The 0.2--1~keV range is directly relevant to the generation of oblique, lower-band chorus, since a field-aligned plateau or weak beam at low parallel velocities suppresses Landau damping and permits growth at large wave-normal angles \cite{Mourenas2015,Artemyev2016,Li2016b,agapitovetal:2015}.
The preferential nightside/premidnight occurrence of the strongest low-energy FAL anisotropies (Figure \ref{fig:occurrences}d, top panel) and their persistence near the magnetic equator ($|\lambda|<10^\circ$; Figure \ref{fig:occurrences}d, bottom panels) is consistent with statistical reports that oblique chorus is preferentially observed close to the equator on the nightside, where freshly injected electrons retain their field-aligned character \cite{agapitovetal:2018,Li2016}, as also depicted in Figures \ref{fig:waves}b,d.

The dawnside region of depressed FAL anisotropy occurrence and enhanced transverse anisotropy identified in Figures \ref{fig:minmaxanisotropy_lo-high_e}j,k and \ref{fig:occurrences}d,e is particularly notable in this context.
Its near mutual exclusivity with the transverse-anisotropy-dominated region, evident on comparing Figures \ref{fig:occurrences}a--c with Figures \ref{fig:occurrences}d--f, is consistent with the reported $\sim$97\% mutual exclusivity of quasi-parallel and oblique chorus emissions in statistical wave surveys \cite{Agapitov2016}, and supports the proposed feedback mechanism whereby oblique chorus, whose generation is supported by low-energy FAL electrons, quenches the transverse anisotropy of the 10--30~keV population and thereby locally suppresses quasi-parallel wave growth \cite{Agapitov2016}.
The transition from a duskside/premidnight bias in FAL anisotropy occurrence (Figure \ref{fig:occurrences}d) to the dawnside/noon bias in transverse anisotropy occurrence (Figure \ref{fig:occurrences}c) reproduces, at the level of the source electron distributions, the reported MLT progression from oblique to quasi-parallel chorus dominance during azimuthal drift and relaxation \cite{Agapitov2016,agapitovetal:2018}.

The enhancement of FAL electron occurrence probabilities toward lower $L$ during elevated $P_{\rm SW}$ and $Kp$ (Figures \ref{fig:minmaxanisotropy_30-200eV}p--r,v--x, \ref{fig:minmaxanisotropy_200-1000eV}p--r,v--x, \ref{fig:anispar_occurrences_psw}, and \ref{fig:anispar_occurrences_kp}), together with the broadened energy extent of these anisotropies noted in Section \ref{sec:stats}, indicates an increased presence of FAL electrons at energies from tens of eV up to $\sim$1~keV during active conditions.
Such electrons fall within the 50--1000~eV range identified as responsible for driving the linear electron-acoustic instability that generates TDS near the geomagnetic equator and in the auroral zone \cite{mozeretal:2015}, and our results therefore predict that the occurrence of TDS-favorable conditions should likewise expand toward lower $L$, and preferentially into the dusk-to-premidnight sector, during enhanced solar wind driving and magnetospheric activity.

\subsection{The Inner Magnetosphere and Plasmasphere}
Within the plasmasphere itself, the very large FAL anisotropies and high occurrence probabilities found at $L\lesssim4$ in the low-energy range (Figures \ref{fig:minmaxanisotropy_lo-high_e}j--l and \ref{fig:occurrences}d) exhibit a pronounced dayside/postnoon bias analogous to the case event distributions in Figure \ref{fig:event1}f, where FAL electrons at energies up to $\sim$100~eV were observed within the plasmasphere on both transects.
This spatial pattern is similar to the dayside enhancement of $\sim$33~eV FAL electron fluxes at $L<2$ reported by \citeA{dentonetal:2017} and attributed to solar-EUV-driven ionospheric outflow, suggesting that a related ionospheric source may sustain the FAL anisotropies observed here at somewhat larger $L$.
Because these low-energy FAL electrons occur at energies below the threshold typically invoked for oblique chorus generation, their principal role may instead be as seed populations for local wave-particle energization processes (e.g., kinetic \alfven\ waves or TDS) rather than as a direct chorus source, an interpretation consistent with the confinement of oblique-chorus-relevant FAL anisotropies to $L\gtrsim6$ in Figures \ref{fig:minmaxanisotropy_lo-high_e}j--l.

\subsection{Perpendicular Anisotropies and Plasmaspheric Hiss}
The high-energy transverse anisotropies found within $L\lesssim 4$ merit separate comment, since they lie in the region where plasmaspheric hiss is generated and sustained.
Figures \ref{fig:minmaxanisotropy_lo-high_e}g--\ref{fig:minmaxanisotropy_lo-high_e}i show that at 1--30~keV the largest transverse anisotropies of the entire survey occur at $L\lesssim 4$, with median $\amin\sim$0.10 (i.e., $\amin^{-1}\sim$10), and Figure \ref{fig:occurrences}c shows that the corresponding occurrence probabilities reach a median of 94\% on the dayside and 83\% on the nightside.
Both are approximately symmetric in MLT, in marked contrast to the low-energy range at the same distances, where transverse anisotropies are weak and nightside-biased.
This identifies a strong, spatially extensive, and MLT-symmetric free-energy reservoir available to whistler-mode hiss throughout the plasmasphere, and is consistent with the broad, MLT-extended hiss amplitude distribution at low $L$ in Figures \ref{fig:waves}a and \ref{fig:waves}c.
It is complementary to, rather than in competition with, the chorus-to-hiss supply mechanism \cite{bortniketal:2008,hartleyetal:2019}: locally generated growth on the anisotropic 1--30~keV population and inward propagation of chorus energy \cite{summersetal:2008,meredithetal:2004} may both contribute, and the near MLT symmetry of the anisotropy reported here provides a testable discriminator, since the chorus-supplied contribution should retain the dawnside preference of its source while the locally supplied contribution should not.

Taken together, the results demonstrate that the transverse and FAL anisotropy populations occupy largely complementary regions of $L$-MLT space whose boundaries shift systematically with solar wind pressure and geomagnetic activity, providing an anisotropy-based framework consistent with, and complementary to, the wave-based statistical picture of chorus and hiss distributions summarized in Figure \ref{fig:waves}.

\section{Summary and Conclusions}
\label{sec:conclusions}

We have performed a statistical survey of the sizes and occurrence probabilities of magnetic field-aligned (FAL) and perpendicular anisotropies in electron distribution functions observed by the THEMIS A and D spacecraft within Earth's near-equatorial magnetosphere at $L\le 14$ between 21 June 2016 and 31 August 2017.
Anisotropies were diagnosed at the level of the distribution function, rather than through temperature moments, using the pair of extremal flux ratios $\amax \equiv {\rm max}(J_{e\parallel}/J_{e\perp})$ and $\amin \equiv {\rm min}(J_{e\parallel}/J_{e\perp})$ evaluated within low (30--200~eV), middle (0.2--1~keV), and high (1--30~keV) energy ranges.
This pairing is central to the present study: it retains the fact that a single distribution can be simultaneously field-aligned at one energy and perpendicular at another, and a location at which both $\amin\gtrsim 1$ and $\amax\gtrsim 1$ unambiguously identifies a region devoid of field-aligned electrons across the full band.
The resulting database of 161,300 spin-resolution measurements was sorted in $L$--MLT and $L$--$\lambda$ and by solar wind dynamic pressure $P_{\rm SW}$ and magnetospheric activity as measured by $Kp$.
Our principal results are as follows.

\begin{enumerate}

\item \emph{Different electron populations carry distinguishable anisotropy signatures.} The case event of 23--24 July 2016 (Figures~\ref{fig:event1} and \ref{fig:edist_event1}) resolves three populations: counterstreaming FAL electrons from just above the spacecraft potential to several hundred eV, with $\amax$ reaching 16--150 (median 42) near the magnetopause; a warm, strongly transverse population at 0.1--1~keV with $\amin^{-1}$ up to $\sim$8; and hot plasma sheet electrons at $\gtrsim$1~keV that are near-isotropic at large $L$ and become increasingly perpendicular with decreasing $L$. The transition from parallel to perpendicular anisotropy with decreasing $L$ arises from a relative enhancement of the transverse flux accompanied by a modest reduction of the field-aligned components, rather than from the disappearance of the field-aligned population alone.

\item \emph{FAL anisotropies are largest and most pervasive in the low energy range}, occurring at essentially all MLT and $L\le 14$. The largest values occur in the dayside/postnoon plasmasphere, where the lower quartile, median, and upper quartile values of $\amax$ are typically 12, 19, and 28, respectively; the nightside plasmasphere is markedly depressed, with occurrence probabilities of $\amax>2$ falling below $\sim$25\% at $L\le 3$. FAL anisotropies and occurrences drop sharply in the trough just outside the plasmasphere before recovering with increasing $L$, and the recovery is more rapid on the duskside, yielding a postnoon/duskside bias in the outer magnetosphere under quiet conditions.

\item \emph{A localized region near dawn at $4\lesssim L\lesssim 8$--9 is typically, though not always, devoid of field-aligned electrons over the entire 30--200~eV band.} This region is identified unambiguously because both $\amin\gtrsim 1$ and $\amax\gtrsim 1$ there, and it appears as a pronounced dropout ($\lesssim$30\%) in the FAL occurrence probability (Figure~\ref{fig:occurrences}d) that broadens in both MLT and $L$ with increasing energy. Its origin is not established here; it may reflect the drift paths that have access to this region, the inaction of the FAL source mechanism there, or both.

\item \emph{FAL anisotropies persist into the middle and high energy ranges but weaken systematically with increasing energy}, and their spatial domain contracts to $L\gtrsim$~8--12 depending on MLT. Their distributions are not symmetric about midnight: the middle energy range forms an asymmetric horseshoe centered on the morning sector, whereas the high energy range forms one centered on premidnight. The systematic change of pattern with energy indicates that more than one field-aligned electron source and/or energization mechanism contributes across the band sampled here.

\item \emph{Perpendicular anisotropies behave in a complementary manner}, being confined closer to Earth than their FAL counterparts and increasing in both magnitude and spatial extent with increasing energy. In the low and middle energy ranges the largest values occupy a broad dayside region at $L\sim$~5--10, but are restricted to $L\sim$~5--8 on the nightside. Occurrence probabilities of $\amin^{-1}>2$ rise from a median of $\sim$80\% at $L\sim$~6--10 on the dayside in the low energy range to $\sim$92\% in the high energy range, where they also extend to the magnetopause in the prenoon sector. Within the plasmasphere, perpendicular anisotropies are weak and nightside-biased at low energies but become strong and approximately MLT-symmetric at 1--30~keV.

\item \emph{The two anisotropy classes are close to mutually exclusive in $L$--MLT space at every energy.} Regions of high FAL occurrence coincide with minima in perpendicular occurrence and vice versa, most clearly in the dawnside dropout described in (3) and in the prenoon transverse-dominated region at $L\gtrsim 6$.

\item \emph{Latitudinal structure differs between the two classes.} Within the $|\lambda|\le 10^\circ$ range accessible to THEMIS A and D, the highest dayside occurrence probabilities of perpendicular anisotropy at 1--30~keV and $L\ge 9$ occur off the magnetic equator, near $|\lambda|\sim$~5--10$^\circ$, whereas the nightside distribution shows no comparable off-equatorial preference. In contrast, the middle-energy FAL anisotropies remain concentrated near $\lambda=0^\circ$ from noon to dusk. These trends are consistent with a picture in which transverse anisotropy is redistributed in latitude during azimuthal drift while field-aligned populations remain equatorially confined near their source, although the restricted latitudinal coverage of these orbits does not by itself establish such a progression.

\item \emph{Both $P_{\rm SW}$ and $Kp$ are important controlling parameters, and $Kp$ exerts the stronger influence.} Under elevated conditions the FAL anisotropy domain expands to lower $L$ at most MLT, with the largest enhancements from postnoon through dusk and near local noon at $L\gtrsim 10$; the perpendicular anisotropy domain shifts inward by roughly $\Delta L\sim 1$ in both the dayside and nightside outer magnetosphere. Simultaneously, the low- and middle-energy perpendicular distributions develop a prenoon bias that is absent under quiet conditions. Under high $Kp$ the low-energy median $\amax$ distribution develops a wedge-like channel of enhanced FAL anisotropy extending from the plasmasphere to $L\sim 14$ near 14--18~MLT that is not present under high $P_{\rm SW}$ alone. This channel lies in the same sector as the enhanced cold plasma densities reported for comparable conditions \cite{walshetal:2020,hulletal:2021} and attributed to plasmaspheric plumes, which is consistent with, though it does not by itself demonstrate, a plume supply of the low-energy field-aligned population in this sector.

\item \emph{The anisotropy distributions provide an observational, source-level counterpart to the statistical wave distributions.} The regions of largest and most probable perpendicular anisotropy at 1--30~keV coincide with the reported source region of quasi-parallel lower-band chorus, and their inward, prenoon-biased confinement under active conditions parallels the corresponding compression of the chorus source region (Figure~\ref{fig:waves}). The strong, MLT-symmetric transverse anisotropies found at 1--30~keV within $L\lesssim 4$ lie in the region where plasmaspheric hiss is generated and sustained, and therefore identify the free-energy reservoir available to that mode \cite{summersetal:2008,meredithetal:2004}. Conversely, the low-energy field-aligned anisotropies, which are strongest from postnoon through premidnight and remain close to the equator, occupy the region in which very oblique chorus is preferentially reported, and their near mutual exclusivity with the transverse regions mirrors the reported mutual exclusivity of quasi-parallel and oblique chorus emissions \cite{Agapitov2016}. The expansion of field-aligned anisotropies at 30~eV--1~keV to lower $L$ and toward dusk under elevated $P_{\rm SW}$ and $Kp$ further implies that conditions favorable to the electron-acoustic generation of Time Domain Structures \cite{mozeretal:2015} should extend inward and duskward during active periods--a prediction that is directly testable with THEMIS and MMS burst-mode wave data.

\end{enumerate}

The results establish that the free-energy sources for the principal wave modes of the inner and outer magnetosphere are organized into well-defined, largely non-overlapping, and systematically activity-dependent regions of $L$--MLT--$\lambda$ space.
Because both the anisotropy magnitude and the plasma density and temperature are available from the same measurements, a natural extension of this work is to convert these distributions into maps of the excess above the marginal-stability threshold for whistler-mode growth \cite{gary+wang:1996}, thereby moving from a description of where free energy resides to a prediction of where it is released.
Comparison of such maps with concurrent measurements of chorus, hiss, and time domain structures--and with kinetic \alfven\ wave Poynting fluxes in the inner magnetosphere \cite{hulletal:2020b}--offers a direct route to quantifying the role of electron anisotropy in magnetospheric energy conversion.

\begin{acknowledgments}
 Work at the University of California, Berkeley was
 supported by NASA grants 80NSSC22K0521, 80NSSC21K1690, 80NSSC23K1278 and NSF Award No 2555054.
We acknowledge NASA contract NAS5-02099 and V. Angelopoulos for use of data from the THEMIS Mission. Specifically, C. W. Carlson and J. P. McFadden for use of ESA data and K. H. Glassmeier, U. Auster and W. Baumjohann for the use of FGM data provided under the lead of the Technical University of Braunschweig and with financial support through the German Ministry for Economy and Technology and the German Center for Aviation and Space (DLR) under contract 50 OC 0302.
\end{acknowledgments}

\section*{Conflict of Interest}
The authors declare no conflicts of interest relevant to this study.

\section*{Open Research}
Level~2 electron electrostatic analyzer (ESA), electric field investigation (EFI), and fluxgate
magnetometer (FGM) data from the THEMIS~A and THEMIS~D spacecraft
\cite{angelopoulos:2008,mcfaddenetal:2008,bonnelletal:2008,austeretal:2008} for the interval
21 June 2016 to 31 August 2017 are publicly available from the THEMIS data archive at the
University of California, Berkeley (https://themis.ssl.berkeley.edu/data/themis/) and from the
NASA Space Physics Data Facility (https://cdaweb.gsfc.nasa.gov/).
Solar wind plasma and magnetic field parameters and geomagnetic activity indices were obtained
from the NASA OMNIWeb service (https://omniweb.gsfc.nasa.gov/).
Data were processed using the publicly available IDL-based Space Physics Environment Data
Analysis System (SPEDAS) \cite{angelopoulosetal:2019}.
The anisotropy database, quartile maps, and occurrence probability maps generated in this study
are derived products of the above archives; no new observations were produced.

\newcommand{\noopsort}[1]{} \newcommand{\singleletter}[1]{#1}

\end{document}